\documentclass[conference]{IEEEtran}

\usepackage{url}
\usepackage[noadjust]{cite}
\usepackage{xspace}
\usepackage{amsmath,amssymb,amstext}
\usepackage{array}
\usepackage{stmaryrd}
\usepackage{enumerate,url,soul}
\usepackage{xargs}
\usepackage{todonotes}
\usepackage[subject={Todo},author={Josef}]{pdfcomment}
\usepackage{balance}
\usepackage{xfrac}
\usepackage{wrapfig}
\usepackage[caption=false,font=footnotesize]{subfig}
\usepackage{multirow}
\usepackage{mwe,tikz}\usepackage[percent]{overpic}
\usepackage[export]{adjustbox}
\usepackage{listings}
\definecolor{keyword}{RGB}{128, 0, 64}
\definecolor{commentgreen}{RGB}{63, 127, 95}
\definecolor{badblue}{RGB}{42,0,255}
\lstdefinelanguage{spec}{
    emph={%  
    begin, action, constraints, end, do, has, value, quantitative, attack, tree, detection, rates, defense, effectiveness, attributes, actions, variables, nodes, countermeasure, model, attackers, init, diagram, attacker, states, transitions, succ,and,format, fail,behavior,allowed,OAND,AND,OR,K,K1,K2,K3,-OR->, remove,seed,steps,file,simulate,exportDTMC,analysis,query,eval,label,with,when,delta,servers,parallelism,alpha,default,ALL,from,to,by%
    },emphstyle={\color{keyword}\bfseries},
    inputencoding=utf8,
    tabsize=2,
    rulecolor=,
    comment=[l][\color{commentgreen}]{//},%
    columns=fixed,
    showstringspaces=false,
    extendedchars=true,
    breaklines=true,
    numbers=none,
    numberfirstline=true,
    frame=single,
    showtabs=false,
    showspaces=false,
    showstringspaces=false,
    escapeinside={(*@}{@*)},
    basicstyle=\fontsize{7}{8}\ttfamily,
    identifierstyle=\fontsize{7}{8}\ttfamily,
    keywordstyle=\fontsize{7}{8}\ttfamily\color[rgb]{0,0,1},
    stringstyle=\fontsize{7}{8}\ttfamily\color[rgb]{0.627,0.126,0.941},
}
\lstdefinelanguage{grammar}{
  basicstyle=\itshape,
  frame=single,
  breaklines=true,
  escapeinside={(*@}{@*)},
  literate={->}{$\rightarrow$}{2}
           {}{$\alpha$}{1}
           {}{$\delta$}{1}
}
\renewcommand{\lstlistingname}{Code}
\newcommand{\lline}[1]{Line~\ref{#1}} % ref line of listings
\newcommand{\llines}[2]{Lines~\ref{#1}-\ref{#2}} % ref lines of listings
\usepackage{float}
\usepackage{stfloats}%\fnbelowfloat

\usepackage{xcolor}

\usepackage{enumitem}

\usepackage{booktabs}

\newcommand{\toolfont}[1]{\textsf{#1}}
\newcommand{\codefont}[1]{\texttt{#1}}
\newcommand{\ourtool}{\toolfont{RisQFLan}\xspace}
\newcommand{\qflan}{\toolfont{QFLan}\xspace}
\newcommand{\conf}{configuration{}}

\usepackage[ligature, inference]{semantic} 
\newcommand{\when}[1]{t(#1)}
\newcommand{\consistent}[1]{\textit{con}(#1)}
\newcommand{\countermeasures}[1]{c(#1)}
\newcommand{\de}[2]{\textit{de}(#1,#2)}
\newcommand{\dr}[1]{\textit{dr}(#1)}

\usepackage{wrapfig}
\newcommand{\mylstinline}[1]{\lstinline[language=spec,mathescape=true,basicstyle=\ttfamily]{#1}}
\newcommand{\lil}[1]{\mylstinline{#1}}
\newcommand{\code}[1]{\lstlistingname~\ref{#1}}

\newcommand{\mv}{MultiVeStA}

\begin{document}

\title{Quantitative Security Risk Modeling and Analysis with RisQFLan}

\author{\IEEEauthorblockN{%
Maurice H. ter Beek\IEEEauthorrefmark{1},
Axel Legay\IEEEauthorrefmark{2},
Alberto Lluch Lafuente\IEEEauthorrefmark{3}, 
Andrea Vandin\IEEEauthorrefmark{4}}
\IEEEauthorblockA{\IEEEauthorrefmark{1}ISTI--CNR, Pisa, Italy}%\\ Email: maurice.terbeek@isti.cnr.it}
\IEEEauthorblockA{\IEEEauthorrefmark{2}UCLouvain, Belgium}%\\ Email: axel.legay@uclouvain.be}
\IEEEauthorblockA{\IEEEauthorrefmark{3}DTU, Lyngby, Denmark}%\\ Email: albl@dtu.dk}
\IEEEauthorblockA{\IEEEauthorrefmark{4}Sant'Anna School of Advanced Studies, Pisa, Italy}}%\\ Email: andrea.vandin@santannapisa.it}}

\maketitle

\begin{abstract}
Domain-specific \emph{quantitative\/} modeling and analysis approaches are fundamental in scenarios in which \emph{qualitative\/} approaches are inappropriate or unfeasible. 
In this paper, we present a tool-supported approach to quantitative graph-based security risk modeling and analysis based on attack-defense trees.
Our approach is based on \qflan, a successful domain-specific approach to support quantitative modeling and analysis of highly configurable systems, whose domain-specific components have been decoupled to facilitate the instantiation of the \qflan approach in the domain of graph-based security risk modeling and analysis. 
Our approach incorporates distinctive features from three popular kinds of attack trees, namely enhanced attack trees, capabilities-based attack trees and attack countermeasure trees, into the domain-specific modeling language.
The result is a new framework, called \ourtool, to support quantitative security risk modeling and analysis based on attack-defense diagrams. 
By offering either exact or statistical verification of probabilistic attack scenarios, \ourtool constitutes a significant novel contribution to the existing toolsets in that domain. 
We validate our approach by highlighting the additional features offered by \ourtool in three illustrative case studies from seminal approaches to graph-based security risk modeling analysis based on attack trees.
\end{abstract}


\section{Introduction}
\label{section:introduction}

\emph{Quantitative} modeling and analysis approaches are essential to support software and system engineering in scenarios where \emph{qualitative} approaches are inappropriate or unfeasible, e.g. due to complexity or uncertainty, or by the quantitative nature of the properties of interest. Automated approaches to support quantitative modeling and analysis have been developed extensively during the last decades, including generic as well as domain-specific approaches (cf., e.g., \cite{BDHHLLT12,KL12,HH15,KRS15,ANP16,BDH16,KS17,BLLV18,BL19b,HHHKKKPQRS19}). 

\qflan~\cite{BLLV18} is one example of a successful domain-specific approach to support quantitative modeling and analysis of highly configurable systems, such as software product lines.
\qflan combines several well-studied rigorous notions and techniques in an Eclipse-based domain-specific tool framework. It consists of a domain-specific language (DSL) tailored for configurable systems, and an analysis engine based on statistical model checking (SMC)~\cite{Agha18,LLTYSG19}. 
In~\cite{BLLV18}, we showed the robustness and scalability of \qflan by analyzing large instances of case studies that could not be analyzed before.

In this paper, we generalize the \qflan approach by decoupling domain-specific components and instantiating the \qflan approach in a new domain: risk modeling and analysis.

The result, called \ourtool, is a new framework to support graph-based \emph{quantitative security risk modeling and analysis}. It constitutes a significant novel contribution to existing toolsets in that domain. 
In particular, \ourtool can be used to: 
\begin{enumerate}
\item build rich models by combining distinctive features from existing formalisms for risk modeling and analysis;
\item enhance the analysis of existing tools for risk modeling.
\end{enumerate}
Regarding~1), the DSL of \ourtool has been designed to include the most significant features of existing formalisms based on attack trees, such that they can be combined in the same model. 
Subsets of the \ourtool DSL, indeed, can thus be used to capture classes of existing modeling formalisms.
In addition, \ourtool allows one to focus on specific dynamic threat profiles, a feature that is being supported only recently by very few approaches (\cite{LWS14,KRS15,ANP16,GHLLO16,HJLLP17,KSRYHBRS18}) and in a limited way (cf.\ the detailed discussion in Section~\ref{section:related}).

We validate feature~2) by showing in Section~\ref{section:validation} how three influential classes of risk models based on attack trees can be specified in \ourtool, and how the \ourtool analysis capabilities can be used to complement and enrich those provided by existing toolsets. This is an advantage offered with respect to the existing tools.
In particular, \ourtool includes an additional analysis engine based on exact probabilistic model checking %~\cite{BK08} 
that is not inherited from \qflan, which comes with a statistical model checking engine~\cite{CHVB18}.

\paragraph*{Synopsis}

Section~\ref{section:risk} introduces the domain of graph-based security risk modeling with attack-defense trees.
Section~\ref{section:qflan} presents a first contribution of the paper: a generalization of the \qflan approach to domain-specific quantitative modeling and analysis. 
Sections~\ref{section:riskflan}-\ref{section:analysis} describe the main contributions of the paper to support security risk modeling and analysis: the \ourtool DSL in Section~\ref{section:riskflan}, its formal semantics in Section~\ref{section:sem} and the analysis capabilities of the \ourtool tool in Section~\ref{section:analysis}.
Section~\ref{section:validation} validates the flexibility of \ourtool by illustrating in detail how features from three influential classes of attack trees can be specified in \ourtool and how the \ourtool analysis capabilities can be successfully used to complement and enrich the analyses provided by existing tools. 
Section~\ref{section:related} discusses related work.
Section~\ref{section:future} draws conclusions and outlines future work.

\section{Graph-based Risk Modeling and Analysis}
\label{section:risk}

This section provides a brief introduction to the specific domain of risk modeling and analysis with graph-based security models. For this purpose, we use as running example the risk assessment of a \emph{``bank robbery''} scenario, which will also be used in Section~\ref{section:riskflan} to illustrate \ourtool. 

Graph-based security models offer an intuitive and effective means to represent security scenarios in complex  systems, by combining intuitive visual features with formal semantics, which can then be used for formal analysis.
\emph{Attack trees} and their variants~\cite{AT,foundationsAT,foundationsADT} constitute a %very 
popular family of graph-based security models for which several approaches have been developed over the last years (cf., e.g., the surveys~\cite{survey,HKCH17,WAFP19}), aiming at providing 
scalable and usable methods for specifying vulnerabilities and countermeasures, their interplay and their key attributes such as cost and effectiveness. Attack trees (and attack-defense trees) thus serve as a basis for quantitative risk assessment, which helps to determine, for instance, where defensive resources are best spent to protect a system. 

\begin{figure}[h]
\centering
\includegraphics[width=\linewidth]{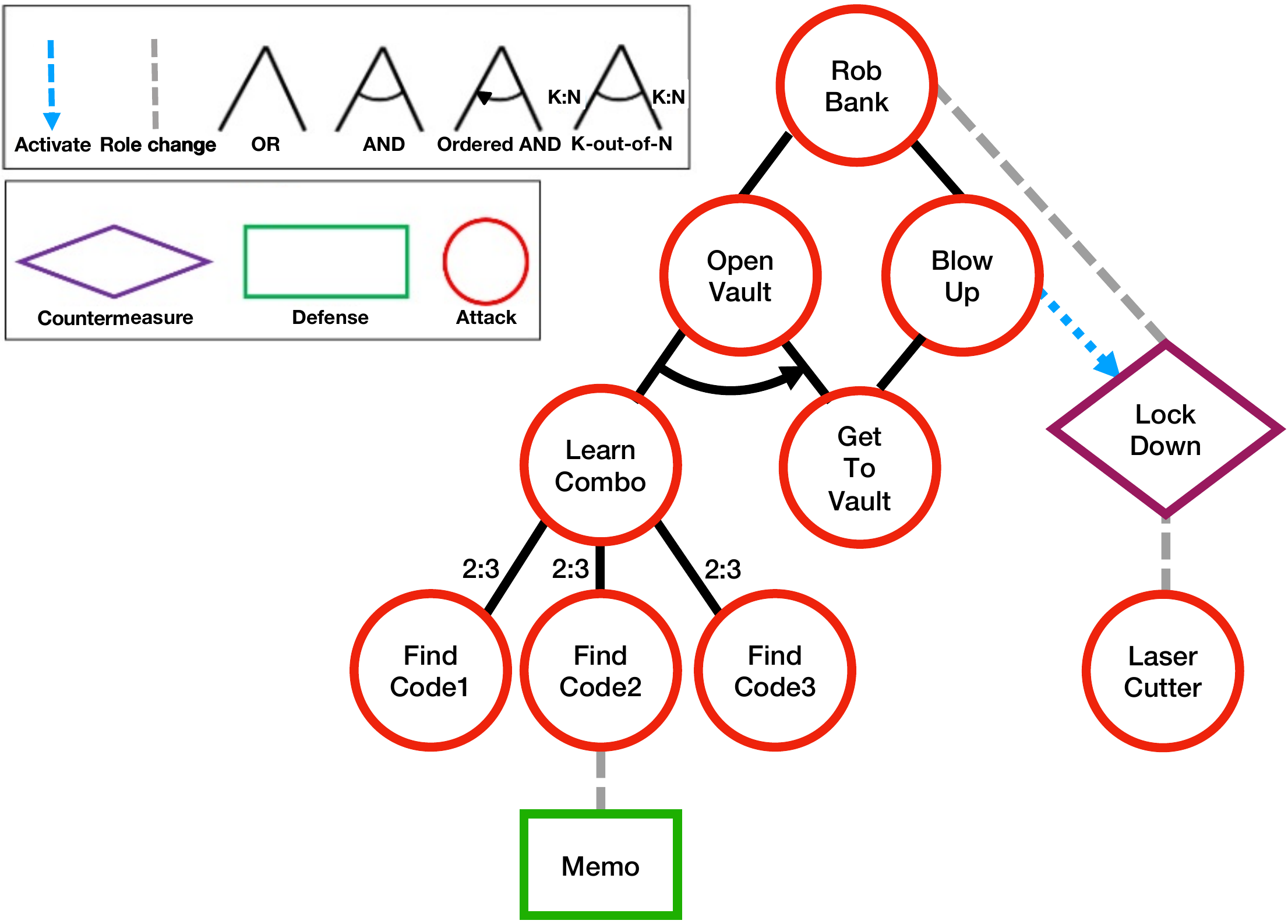}
\vspace*{-0.25cm}
\caption{\label{fig:structure}Attack-defense diagram}
\end{figure}

In their simplest form, attack-defense diagrams are and/or-trees whose nodes represent either attack goals or defensive measures, and with sub-trees representing refinements of such goals and measures. 
Fig.~\ref{fig:structure} shows an attack-defense diagram modeling our running example. The tree's root represents the main threat under analysis, i.e.\ robbing a bank (\codefont{Rob\-Bank}). 

Attack nodes can be refined in several ways by identifying necessary sub-goals and combining them in different ways, e.g.\ with disjunction, (ordered) conjunction, %ordered conjunction, 
etc. 
In our example, the attacker has two options to achieve its main goal: either open the vault (\codefont{Open\-Vault}) \emph{or\/} blow it up (\codefont{Blow\-Up}). This is specified in the tree with corresponding nodes as children of node \codefont{Rob\-Bank}, combined in a disjunctive way.
Another kind of refinement illustrated in our example is the following: in order to open the vault (\codefont{Open\-Vault}), the attacker needs to \emph{first\/} learn its combo (\codefont{Learn\-Combo}) \emph{and\/} \emph{then\/} get to the vault (\codefont{Get\-To\-Vault}). This is specified by combining \codefont{Learn\-Combo} and \codefont{Get\-To\-Vault} through an ordered conjunction.  
A last example of refinement is used to model that for security reasons \emph{two out of three\/} of the vault's opening codes are required (\codefont{Find\-Code1}--\codefont{Find\-Code3}). Instead, blowing it up only requires to get to the vault. 

Attack-defense diagrams can also include defensive mechanisms to deal with or to prevent attack threats. In the example scenario, there are two defensive mechanisms. First, a \codefont{Lock\-Down} \emph{countermeasure}, triggered by (successful or not) blow up attacks that, once active, mitigates bank robbery attacks. The rationale is that the vault is sealed to prevent robbery when an explosion is detected. The second defensive measure in our running example is a \emph{defense\/} \codefont{Memo}, permanently active against attacks trying to find opening code~2 (\codefont{Find\-Code2}). The interplay between such a defensive countermeasure and the corresponding attack nodes is also typically depicted visually, as in our example. Defensive mechanisms, in turn, can also be affected (e.g.\ disabled or mitigated) by attacks. For instance, in our example an attack with a \codefont{Laser\-Cutter} can break the \codefont{Lock\-Down}.

Attack-defense diagrams, besides being a useful tool for modeling and informally reasoning on security risk scenarios, often also have a formal meaning that lies at the basis of formal reasoning, typically supported by effective software tools like Secur\emph{IT}ree~\cite{securitree}, ADTool~\cite{ADTool}, SPTool~\cite{KKB16}, and ATTop~\cite{KSRYHBRS18} to mention a few (cf.~surveys~\cite{survey,WAFP19,HKCH17} for further examples).

Standard analyses conducted on attack-defense diagrams typically regard the feasibility of attacks (e.g.\ \emph{can the attacker activate some actions that will result in the achievement of her/his main goal?}), their likelihood (e.g.\ \emph{what is the probability that the main goal is achieved?}) or their cost (e.g.\ \emph{what is the cheapest successful attack for the attacker?}). Analysis techniques are often based on constraint solving, optimization and statistical techniques. Section~\ref{section:analysis} will provide some of these analyses applied to our running example.

\section{Generalizing the \qflan Approach}
%\section{Generalizing the QFl\lowercase{an} approach to Domain-Specific Quantitative Modeling and Analysis}
\label{section:qflan}

This section describes how the \qflan architecture was made amenable for instantiation in domains beyond the one for which it was %originally 
conceived (configurable systems like software product lines), and how its analysis capabilities were enriched. %, and we outline a methodology to implement new instantiations.

\paragraph*{The original \qflan architecture}

%\begin{figure}[b]
%  \centering
%  \includegraphics[width=0.675\linewidth]{figures/architecture-tse}
%  \vspace*{-0.15cm}
%  \caption{\label{figure:tsearchitecture}The \qflan architecture reproduced from~\cite{BLLV18}}
%\end{figure}

We first summarize the original \qflan architecture as presented in~\cite{BLLV18}, %Fig.~\ref{figure:tsearchitecture} provides a high-level overview of the architecture, which is 
organized in two layers: the Graphical User Interface (\codefont{GUI)}, devoted to modeling, and the \codefont{CORE} layer, devoted to analysis. Both layers are wrapped in an Eclipse-based tool embedding the third-party statistical analyzer \mv{}~\cite{SV13,GRV17}. \qflan is an open-source tool.  
The components of the \codefont{GUI} layer are: 
\begin{itemize}[partopsep=-4pt,topsep=2pt,parsep=4pt,itemsep=-4pt]
\item a \codefont{QFLAN Editor} with editing 
support that is typical of a modern Integrated Development Environment (IDE), developed in the \codefont{XTEXT} framework, and a \codefont{MultiQuaTEx Editor} for property specification in the MultiQuaTEx language~\cite{SV13};
\item a set of \codefont{Views}, including a project explorer, a diagnosis console and a plot viewer for displaying analysis results.
\end{itemize}
The components of the \codefont{CORE} layer are:
\begin{itemize}[partopsep=-4pt,topsep=2pt,parsep=4pt,itemsep=-4pt]
\item a \codefont{Probabilistic Simulator}, which is an interpreter of the formal semantics as probabilistic processes. This interacts with the external statistical analyzer \mv{} to obtain %statistical model checking 
SMC capabilities;
\item a \codefont{Built-in Constraint Solver} used by the simulator to check constraints during simulation.
\end{itemize}

\paragraph*{The refactored \qflan architecture}

\begin{figure}[h]
  \centering
  \includegraphics[width=\linewidth]{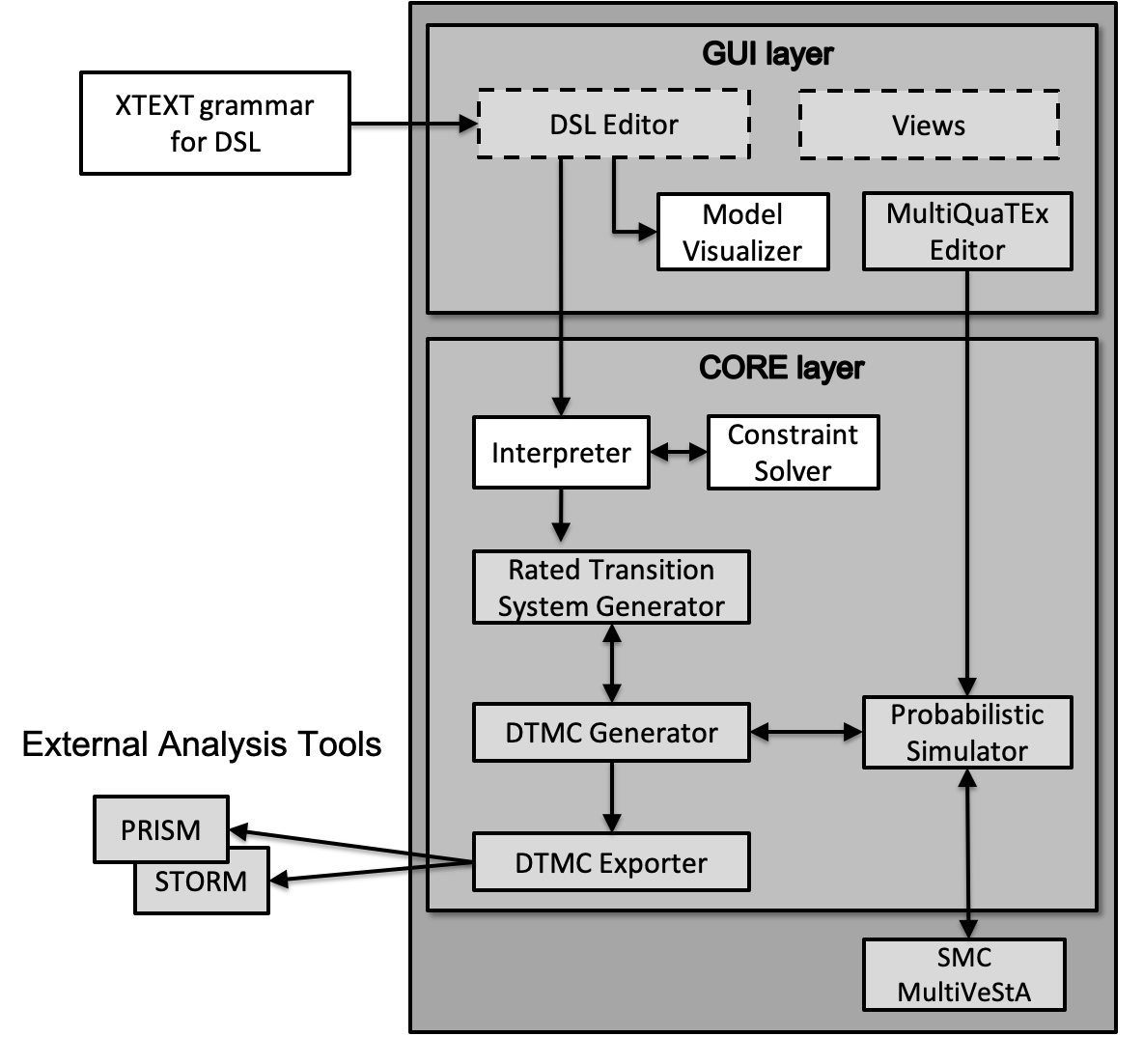}
  \vspace*{-0.5cm}
  \caption{\label{fig:QFLAN}The refactored \qflan architecture}
\end{figure}

The architecture illustrated in Fig.~\ref{fig:QFLAN} decouples domain-specific components of the \qflan architecture from domain-generic ones. The domain-specific components that need to be provided to instantiate the architecture in a new domain are the following: the \codefont{XTEXT grammar for DSL}, the  \codefont{Interpreter}, the \codefont{Constraint Solver} and the \codefont{Model Visualizer} (differentiated %in Fig.~\ref{fig:QFLAN} 
from other components by their blank background). The remaining components are either existing domain-generic components (solid border) or domain-specific components, automatically generated by XTEXT (dashed border). 

The \codefont{GUI} layer basically remains unchanged, except that Fig.~\ref{fig:QFLAN} makes explicit that the \codefont{DSL Editor} is generated automatically from an \codefont{XTEXT grammar for DSL}.
Moreover, it was extended with a \codefont{Model Visualizer} component to offer an automatic visual representation of the model at hand. This is obtained  by %just 
providing an encoding of the models' features of interest in the DOT language\,\footnote{\url{http://www.graphviz.org/doc/info/lang.html}}.

The main changes in the refactored \qflan architecture however concern the \codefont{CORE} layer, whose new components are:
\begin{itemize}[partopsep=-4pt,topsep=2pt,parsep=4pt,itemsep=-4pt]
\item an \codefont{Interpreter} and a \codefont{Constraint Solver}, implementing the formal semantics of the DSL based on rated transition systems (transition systems with rate-decorated transitions);
\item a \codefont{Rated Transition System Generator} relying on the \codefont{In\-ter\-preter} to generate rated transition systems on-the-fly;
\item a \codefont{DTMC Generator}, which uses the \codefont{Rated Transition Sys\-tem Generator} to normalize rated transition systems into on-the-fly generated Discrete-Time Markov Chains (DTMC);
\item a \codefont{Probabilistic Simulator}, which is now separated from the above components and which is able to simulate a DTMC without fully generating it using the on-the-fly \codefont{DTMC Generator}; 
\item a \codefont{DTMC Exporter} which generates an entire DTMC by using the \codefont{DTMC Generator} and exports it in the input format of the well-known probabilistic model checkers PRISM~\cite{KNP11} (\url{www.prismmodelchecker.org}) and STORM~\cite{DJKV17} (\url{www.stormchecker.org}).
\end{itemize}

\section{\ourtool{} DSL%: an Instantiation of \qflan
}
\label{section:riskflan}

\begin{figure*}[t]
  \centering
  \includegraphics[width=\linewidth]{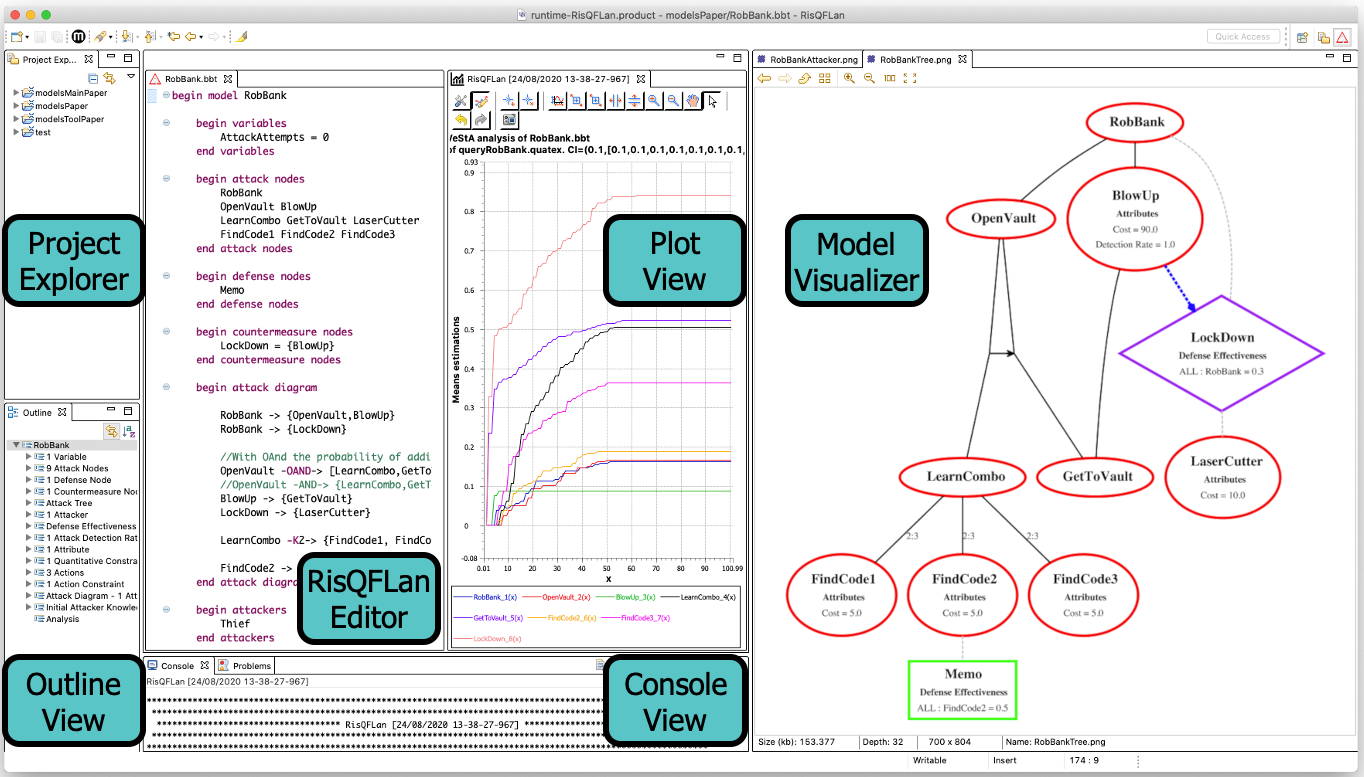}
  \vspace*{-0.25cm}
  \caption{\label{figure:screenshot}A screenshot of \ourtool}
\end{figure*}

This section describes \ourtool, a domain-specific instantiation of \qflan in the security risk domain described in Section~\ref{section:risk}.
A screenshot of \ourtool is provided in Fig.~\ref{figure:screenshot}, depicting the implemented components from the GUI layer in Fig.~\ref{fig:QFLAN}.  
We describe here the DSL of \ourtool, while its formal semantics is given in Section~\ref{section:sem} and its analysis capabilities are presented in Section~\ref{section:analysis}. 
%\subsection{RisQFLan DSL}
\label{section:dsl}
We illustrate the DSL of \ourtool through the running example, whose attack-defense diagram is depicted in Fig.~\ref{fig:structure} (and in Fig.~\ref{figure:screenshot}). 

In the DSL, nodes are declared in specific blocks, cf.\,\code{spec:nodes}. 
Note that countermeasure nodes require to indicate the attack node(s) that may trigger them.

\begin{wrapfigure}{r}{0.44\linewidth}
\vspace{-0.5cm}
\begin{lstlisting}[language=spec,caption={Nodes},captionpos=b, label={spec:nodes},numbers=none,escapeinside={@}{@}]
begin attack nodes
 RobBank OpenVault BlowUp
 LearnCombo GetToVault 
 FindCode1 FindCode2
 FindCode3 LaserCutter
end attack nodes

begin defense nodes
 Memo
end defense nodes

begin countermeasure nodes
 LockDown = {BlowUp}
end countermeasure nodes
\end{lstlisting}
\vspace{-0.25cm}
\end{wrapfigure}

Our attack-defense diagrams relate nodes by two types of relations: (i)~\emph{refinements\/} shape offensive (defensive, resp.) nodes into a set of offensive (defensive, resp.) sub-nodes; (ii)~\emph{role-changes\/} state how to oppose offensive (defensive, resp.) nodes by defensive (offensive, resp.) nodes. Each node has at most one refinement and at most one role-change. 
Typical for our approach is that nodes may have multiple parents, which is convenient to specify an attack (defense) node that affects multiple defenses (attacks) or an attack (defense) node that refines many attacks (countermeasures).

We offer \lil{OR}, \lil{AND}, \lil{OAND} (ordered \lil{AND}), and \codefont{k-out-} \codefont{of-n} refinements for attack and countermeasure nodes. Defense nodes model static, atomic defenses that cannot be refined. Countermeasures are also atomic, but they can be refined with defense nodes to permit \emph{reactive\/} defense nodes that become effective only upon (attack detection and) activation of the refined countermeasure.
\lil{AND} and \lil{OR} refinements originate from the seminal works on attack trees~\cite{AT}. \lil{OAND} refinements stem from \emph{enhanced} %~\cite{CY07} 
and \emph{improved attack trees}~\cite{CY07,improved} and are used to model ordered attacks: sub-nodes can be activated in any order but only the correct order activates the parent node. 
% commented out for brevity. It explains a design choice between the two above papers.
%. adding a sequential aspect has been  in two ways: the children can be activated either in a given order only or in any order, but only the correct order satisfies the relation. We adopt the latter, since orders of attacks can be modeled as attack behavior, and it is interesting to model an attacker that does not know the correct attack order. 
The \codefont{k-out-of-n} refinements are inspired by \emph{attack countermeascure trees}~\cite{act}.  

\llines{hier1}{hier2} of \code{spec:hcon} show how to declare attack diagrams in \ourtool. The square brackets of \lil{OAND} indicate that order matters: \codefont{Open}\-\codefont{Vault} requires \codefont{Learn}\-\codefont{Combo} and \codefont{GetTo}\-\codefont{Vault} \emph{in that order}. \lil{K2} expresses that \emph{at least two\/} of the three sub-attacks of \codefont{Learn}\-\codefont{Combo} are required. 
Inspired by other formalisms supporting both attack and defense mechanisms, like \emph{attack-defense trees}~\cite{foundationsADT}, a \textit{role-changing\/} relation describes the attack a countermeasure or defense works against (e.g.\ \codefont{LockDown} defends against \codefont{RobBank}) or vice versa (e.g.\ \codefont{LaserCutter} neutralizes \codefont{LockDown}).  
\llines{change1}{change2} of \code{spec:hcon} show that attack, defense and countermeasure nodes can additionally have a \emph{role-changing\/} relation with a child of the opposite role, an opponent node affecting its activation.

\begin{lstlisting}[language=spec,caption={Attack-defense diagram},captionpos=b, label={spec:hcon},numbers=left,escapeinside={@}{@}]
begin attack diagram
 RobBank -> {OpenVault, BlowUp} @\label{hier1}@
 OpenVault -OAND-> [LearnCombo, GetToVault] 
 BlowUp -> {GetToVault} 
 LearnCombo -K2-> {FindCode1, FindCode2, FindCode3} @\label{hier2}@
 RobBank -> {LockDown} @\label{change1}@
 LockDown -> {LaserCutter}
 FindCode2 -> {Memo} @\label{change2}@
end attack diagram
\end{lstlisting}

As in other approaches~\cite{survey}, attack nodes may be decorated with attributes, like cost or detection rates, for quantitative analyses~\cite{KMS12,ANP16,HJLLP17}. The cost of (attempting) an attack, like the attribute \codefont{Cost} in \code{spec:att}, may be used to impose constraints. The default value is~$0$, e.g.\ $\codefont{Cost(GetToVault)}\!=\!0$.
The cumulative value for the entire scenario, often the cost associated to a (sub-system rooted in a) node, is the sum of the costs of its active descendants~\cite{AT}. However, the total cost of an attack should not reflect only the cost of \emph{successful\/} sub-attacks, as this would be a best-case scenario. Therefore, in \ourtool we consider both successful and failed attack attempts to compute the value of an attribute of an attack node. Furthermore, we allow attributes also for defensive nodes. 

\begin{lstlisting}[language=spec,caption={Attributes},captionpos=b, label={spec:att},numbers=none]
begin attributes
 Cost = {LaserCutter = 10, BlowUp = 90, 
         FindCode1 = 5, FindCode2 = 5, FindCode3 = 5}
end attributes
\end{lstlisting}

In~\cite{securitree,whitepaper}, a \emph{noticeability\/} attribute is a behavioral metric used to indicate the likeliness of an attack attempt to be noticed. Following \emph{attack countermeasure trees}~\cite{act}, we make this notion a first-class citizen of \ourtool, called \emph{attack detection rate\/}, which influences activation of countermeasures. More precisely, such a rate determines the
\begin{wrapfigure}{r}{0.48\linewidth}
\vspace{-0.2cm}
\begin{lstlisting}[language=spec,caption={Attack detection rates},captionpos=b, label={spec:dr},numbers=none]
begin attack detection rates
 BlowUp = 1.0
end attack detection rates
\end{lstlisting}
\vspace{-0.2cm}
\end{wrapfigure}
probability for an attack attempt, whether successful or not, to be detected, and it triggers the activation of the affected countermeasures, in the sense that higher detection rates lead to more likely activation of countermeasures. The default value is~$0$, i.e.\ an attack is undetectable. \code{spec:dr} shows that an attempt to blow up a vault is always noticed.

In~\cite{foundationsADT,ADTool}, an attack node is \emph{disabled\/} if it is affected by a defense. However, a common conception in security is that nothing is 100\% secure. Therefore, we include the notion of \textit{defense effectiveness\/} from~\cite{act} to specify the probability for a defense node to be effective against a combination of attack nodes and attack behavior. The rationale is that different attackers might be affected differently, even when attempting the same attack (e.g.\ a security guard is efficient against a thief, but not against a military attack). The default value is $0$, i.e.\ the defense has no effect. \code{spec:de} states that \codefont{Memo} scales the probability of succeeding in \codefont{FindCode2} attacks by $1 - 0.5$, whereas \codefont{LockDown} scales that of \codefont{RobBank} by $1 - 0.3$.

\begin{lstlisting}[language=spec,caption={Defense effectiveness (\codefont{ALL} denotes any attacker)},captionpos=b, label={spec:de},numbers=none]
begin defense effectiveness
 Memo(ALL, FindCode2) = 0.5, LockDown(ALL, RobBank) = 0.3
end defense effectiveness
\end{lstlisting}

\subsection{Attack Behavior}

An important feature of our models is that defensive behavior is \emph{reactive\/}, while attackers are \emph{proactive}.
\ourtool allows to fine tune security scenarios by defining explicit \emph{attack behavior}, implicitly constrained by an attack-defense diagram. 
The combination of attack-defense diagrams and explicit (probabilistic) attack behavior was motivated by work on configurable systems~\cite{BLLV18,VBLL18}. 
Explicit attack behavior enables the analyses of specific attacker types, like script kiddies, insiders, and hackers, which has the advantage of being able to evaluate system vulnerabilities for those attacker types that make more sense for the security scenario at hand. Moreover, it enables novel types of analysis to complement the classical best- and worst-case evaluations of attack graphs (like the bottom-up evaluation in ADTool~\cite{ADTool}). 

\begin{figure}[h]%{r}{0.58\linewidth}
%\centering
%\vspace{-0.2cm}
%\hspace{-0.25cm}
\includegraphics[width=\columnwidth]{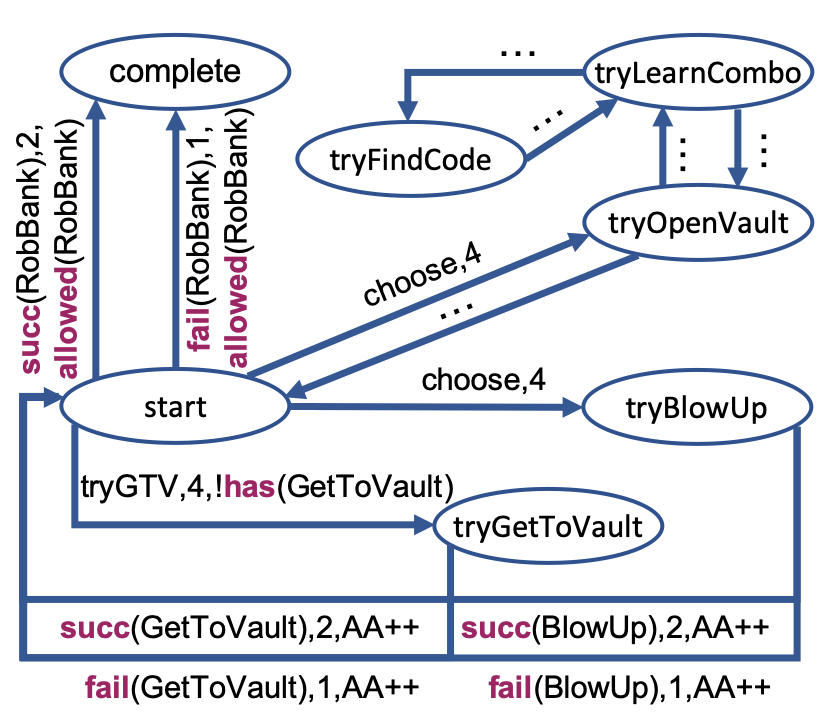}
\vspace{-0.5cm}
\caption{\label{fig:behavior}Attack behavior}   
%\vspace{-0.25cm}
\end{figure}

Attack behavior is modeled as rated transition systems, whose transitions are labeled with the action being executed and a rate (used to compute the probability of executing the action), and possibly with effects (updates of variables) and guards (conditions on the action's executability), in this order (e.g.\ \llines{ls:order1}{ls:order2} in \code{spec:attackers:behavior}). 
Fig.~\ref{fig:behavior} (and its corresponding \code{spec:attackers:behavior}) sketches an attacker, named \codefont{Thief}, that \codefont{start}s by choosing to attempt an open vault (\codefont{try}\-\codefont{Open}\-\codefont{Vault}) or blow up (\codefont{tryBlowUp}) attack.  
Independently of this choice, s/he can try get-to-vault attacks (\codefont{tryGetToVault}), 
required by both strategies. 
\codefont{OpenVault} requires to try to learn the combo, %a \codefont{LearnCombo} attack, %attempted in \codefont{tryLearnCombo}, 
which in turn requires to try to find at least two codes. %, attempted in \codefont{tryFindCode}. 

\begin{wrapfigure}{r}{0.3\linewidth}
\vspace{-0.2cm}
\begin{lstlisting}[language=spec,caption={Actions},captionpos=b, label={spec:actions},numbers=none]
begin actions
 choose tryGTV try
end actions
\end{lstlisting}
\vspace{-0.2cm}
\end{wrapfigure}

Attacker actions can be \emph{user-defined\/} for scenario-specific behavior not directly related to node activation, such as %\codefont{choose}, \codefont{tryGTV} and \codefont{try}.
those in \code{spec:actions} (where \codefont{try} is part of the attacks \codefont{try}\-\codefont{Open}\-\codefont{Vault}, \codefont{try}\-\codefont{Learn}\-\codefont{Combo} and \codefont{try}\-\codefont{Find}\-\codefont{Code} that are not further detailed in Fig.~\ref{fig:behavior} and \code{spec:attackers:behavior}). 
\ourtool also provides a set of predefined attacker actions, like \lil{succ} and \lil{fail} for a successful or failed attack \emph{attempt}, resp., modeled by a probabilistic choice between \lil{succ} and \lil{fail} actions, whose associated rates determine the success likelihood together with (the effectiveness of) the involved defenses. In Section~\ref{section:validation}, we will see how attackers can apply backtracking strategies via the predefined action \lil{remove}.

Attack behavior is executed by considering, at each step, the outgoing transitions from the current state admitted by the attack diagram and by further constraints discussed below. 
Normalizing the sum of the rates of these transitions to 1 leads to a DTMC, while probabilistic simulations are obtained by selecting one transition probabilistically using the transition rates (e.g., from \codefont{start} to \codefont{complete} with probability $\frac{1}{1+2}$). 
%
%Attack behavior can also model attempts of an attacker to obtain information about the scenario (like whether an attack succeeded) by using \codefont{query} actions. Finally, a succeeded attack can be removed with \codefont{remove}, which may turn out useful in the presence of \lil{OAND} relations, since it allows an attacker to apply backtracking strategies to try different attack sequences.
%

Transitions can contain \emph{guards}, like \lil{allowed}, used to attempt \codefont{RobBank} in \codefont{start} only if all required sub-attacks succeeded (cf.\ \llines{ls:start}{ls:start2} in \code{spec:attackers:behavior}), or \lil{!has}, used to forbid the transition to \codefont{try}\-\codefont{Get}\-\codefont{To}\-\codefont{Vault} if one already succeeded to \codefont{Get}\-\codefont{To}\-\codefont{Vault} (cf.\ \lline{ls:vault} in \code{spec:attackers:behavior}). 

\begin{lstlisting}[float=*,language=spec,caption={Attack behavior},captionpos=b, label={spec:attackers:behavior},numbers=left,escapeinside={@}{@}]
begin attacker behavior@\label{ls:attbhv}@
 begin attack
  attacker = Thief
  states = start, tryOpenVault, tryLearnCombo, tryFindCombo, tryGetToVault, tryBlowUp, complete @\label{ls:attstates}@ 
  transitions = 
   start -(succ(RobBank), 2, allowed(RobBank)) -> complete, //If I open or blow up the vault, then I can rob the bank @\label{ls:start}@ 
   start -(fail(RobBank), 1, allowed(RobBank)) -> complete, @\label{ls:start2}@
   start -(tryGTV, 4, !has(GetToVault)) -> tryGetToVault, //Whatever strategy was used, I must get to the vault @\label{ls:vault}@
    tryGetToVault -(succ(GetToVault), 2, {AttackAttempts=AttackAttempts+1}) -> start,
    tryGetToVault -(fail(GetToVault), 1, {AttackAttempts=AttackAttempts+1}) -> start,@\label{ls:vault2}@
    start -(choose, 4) -> tryOpenVault, //This is the strategy where I open the vault
    tryOpenVault -(succ(OpenVault), 2, {AttackAttempts=AttackAttempts+1}, has(LearnCombo) and has(GetToVault)) -> start, @\label{ls:order1}@
    tryOpenVault -(fail(OpenVault), 2, {AttackAttempts=AttackAttempts+1}, has(LearnCombo) and has(GetToVault)) -> start, @\label{ls:order2}@
    tryOpenVault -(try, 2, has(LearnCombo) and !has(GetToVault)) -> start, //I know the combo but did not get to the vault
    tryOpenVault -(try, 5, !has(LearnCombo)) -> tryLearnCombo,  
     ... //Similar for tryLearnCombo and then tryFindCode
   start -(choose, 4) -> tryBlowUp, //This is the strategy where I blow up the vault
   tryBlowUp -(succ(BlowUp), 2, {AttackAttempts=AttackAttempts+1}) -> start,
   tryBlowUp -(fail(BlowUp), 1, {AttackAttempts=AttackAttempts+1}) -> start
 end attack
end attacker behavior@\label{ls:attbhv2}@
\end{lstlisting}

%   // I don't try to open the vault if I didn't succeed in learning the vault and getting to the vault
%   tryOpenVault -(succ(OpenVault), 2, {AttackAttempts=AttackAttempts+1}, has(LearnCombo) and has(GetToVault)) -> start, 
%   tryOpenVault -(fail(OpenVault), 1, {AttackAttempts=AttackAttempts+1}, has(LearnCombo) and has(GetToVault)) -> start,
%   // I managed to learn the combo, but I still did not get to the vault
%   tryOpenVault -(try, 2, has(LearnCombo) and !has(GetToVault)) -> start,
%   tryOpenVault -(try, 5, !has(LearnCombo)) -> tryLearnCombo,
%   // I don't know how many combos I need (I don't use the guard allowed (LearnCombo))
%   tryLearnCombo -(succ(LearnCombo), 5, {AttackAttempts=AttackAttempts+1}) -> tryOpenVault,
%   tryLearnCombo -(fail(LearnCombo), 1, {AttackAttempts=AttackAttempts+1}) -> tryOpenVault,
%   // I don't know how many combos I need to find (2 or 3)
%   tryLearnCombo -(try, 5, !has(FindCode2) or !has(FindCode3)) -> TryFindCombo,
%   tryFindCombo -(succ(FindCode2), 1, {AttackAttempts=AttackAttempts+1}, !has(FindCode2)) -> tryLearnCombo,
%   tryFindCombo -(fail(FindCode2), 5, {AttackAttempts=AttackAttempts+1}, !has(FindCode2)) -> tryLearnCombo,
%   tryFindCombo -(succ(FindCode3), 1, {AttackAttempts=AttackAttempts+1}, !has(FindCode3)) -> tryLearnCombo,
%   tryFindCombo -(fail(FindCode3), 5, {AttackAttempts=AttackAttempts+1}, !has(FindCode3)) -> tryLearnCombo,

%		//LockDown cannot be activated if we have LaserCutter (which is its role-changing son)
%		//Thief = {FindCode1,LaserCutter}

\ourtool also supports \emph{action constraints}, acting as guards on any transition executing a given action (while transition guards constrain single transitions). They are given as \lil{do}$(\textit{act\/}) \rightarrow b$, where $\textit{act\/}$ is an action and $b$ is a Boolean expression over attributes.
As defined in \code{spec:acon}, any transition with action \codefont{choose} is disabled as soon as one succeeds to open or blow up the vault. %This is because one needs to succeed only in one attack strategy at a time. 

%\begin{wrapfigure}{r}{0.44\linewidth}
\begin{lstlisting}[language=spec,caption={Action constraints},captionpos=b, label={spec:acon},numbers=none]
begin action constraints
 do(choose) -> !(has(OpenVault) or has(BlowUp))
end action constraints
\end{lstlisting}
%\vspace{-0.3cm}\end{wrapfigure}

Transitions can also be labeled with \emph{side-effects}: real-valued variables updated upon a transition's execution. Variables model context information, thus allowing for rich descriptions of system states, of attackers and of defenses, greatly facilitating the expression of constraints and the analysis phase. 
\code{spec:var} defines variable \codefont{Attack}\-\codefont{Attempts} (\codefont{AA} in Fig.~\ref{fig:behavior}), which stores the number of attack attempts, updated each time a \lil{succ} or \lil{fail} action occurs as attempt to rob the bank.

\noindent\begin{minipage}{\columnwidth}
%\centering
\begin{minipage}[t]{0.32\linewidth}
%\begin{wrapfigure}{r}{0.34\linewidth}
\begin{lstlisting}[language=spec,caption={Variables},captionpos=b, label={spec:var},numbers=none,escapeinside={@}{@},mathescape=true]
begin variables
 AttackAttempts = 0
end variables
\end{lstlisting}
%\vspace{-0.25cm}
%\end{wrapfigure}
\end{minipage}
\qquad\quad\,\ 
%\centering
\begin{minipage}[t]{0.52\linewidth}
%\begin{wrapfigure}{r}{.95\linewidth}
\begin{lstlisting}[language=spec,caption={Quantitative constraints},captionpos=b, label={spec:qcon},numbers=none]
begin quantitative constraints
 {value(Cost) <= 100}
end quantitative constraints
\end{lstlisting}
%\vspace{-0.25cm}
%\end{wrapfigure}
\end{minipage}
\end{minipage}

In addition to constraints imposed by attack diagrams, transition guards and action constraints, attack behavior may be constrained by quantitative constraints in the form of Boolean expressions involving (arithmetic expressions or inequalities over) reals, attributes and variables.
In \code{spec:qcon}, we constrain to 100 the maximum accumulated cost of an attack, of particular interest since attack behavior may model failed attacks. 

Attack behavior is completed with an initial setup specifying the attacker and any initially accomplished attack(s). The latter
\begin{wrapfigure}{r}{0.34\linewidth}
\vspace{-0.22cm}
\begin{lstlisting}[language=spec,caption={Initial setup},captionpos=b, label={spec:init},numbers=none,escapeinside={@}{@},mathescape=true]
begin init @\label{ls:attinit}@
 Thief = {FindCode1}
end init @\label{ls:attinit2}@
\end{lstlisting}
\vspace{-0.24cm}
\end{wrapfigure}
enrich expressiveness, since one can assign an initial advantage to attackers: an attack-defense diagram models all possible attacks, but some attackers (e.g., insiders) may already have access to critical components. This is convenient as a diagram's sub-trees may be ignored without their explicit removal. Due to \code{spec:init}, the attacker \codefont{Thief} already has one code.

Note that \ourtool provides a programming-like environment that may be attractive to software developers, but it integrates at the same time a graphical component shown in Fig.~\ref{figure:screenshot}, which may make it more attractive for security experts. The DSL moreover has a formal semantics, defined next.

\section{\ourtool{} Operational Semantics}
\label{section:sem}
\subsection{\ourtool{} Models and Configurations}
%\begin{definition}[\ourtool{} model]\label{def:model}
In this section, we provide a formal definition of the ingredients composing \ourtool{} models. In order to improve readability, we provide references to the corresponding code blocks from Section~\ref{section:dsl} when relevant, which show ho the components of the model are actually specified in our DSL. 

	A \ourtool{} model $\mathcal{S}$ is defined as a septuple $\mathcal{S} = \langle \mathcal{N},\mathcal{D},\mathcal{V},\mathcal{A}, \mathcal{B},\mathcal{C},\mathcal{P} \rangle$, where
	\begin{itemize}
		\item 
		$\mathcal{N} =  \mathcal{N}_a \uplus \mathcal{N}_d \uplus \mathcal{N}_c$ is a set of nodes divided into %finite 
		 attack nodes~$\mathcal{N}_a$, defense nodes~$\mathcal{N}_d$ and countermeasure nodes~$\mathcal{N}_c$ (\code{spec:nodes});
		\item 
		$\mathcal{D}$ is a %finite 
		set of attacker actions. The set $\mathcal{D}$ contains all actions  \lil{succ}($n_a$), \lil{fail}($n_a$), and \lil{remove}($n_a$), %and \lil{query}($n$), 
		where $n_a \in \mathcal{N}_a$, and additionally   user-defined actions (\code{spec:actions}); 
			\item 
		$\mathcal{V}$ is a %finite 
		set of variables (\code{spec:var}); 
		\item 
		$\mathcal{A}$ is a %finite 
		set of attackers names (\code{spec:attackers:behavior});
		\item 
		$\mathcal{B}$ is a set of attacker behaviors (\code{spec:attackers:behavior}); 
		\item 
		$\mathcal{C}$ is a set of constraints on the (presence/absence of) nodes, their attributes, and on  (user-defined) variables. Such constraints are formed by the  %finite 
		hierarchical constraints (built with \mbox{\lil{-OR->}}, \mbox{\lil{-AND->}}, \mbox{\lil{-OAND->}}, and \mbox{\lil{-K}{\color{keyword}\bfseries\itshape n}\lil{->}}, \code{spec:hcon}), 
		action constraints (of the form \lil{do}$(\textit{act\/}) \rightarrow b$, where $\textit{act\/}\in \mathcal{D}$ and $b$ is a Boolean expression over attributes, \code{spec:acon}) 
		and quantitative constraints (Boolean expressions enriched with special attributes like \lil{allowed}($n_a$) and \lil{has}($n_a$), with $n_a\in \mathcal{N}_a$, \code{spec:qcon}); 
		\item 
		$\mathcal{P}:  \mathcal{N} \rightarrow \mathbb{R}$ is a %finite 
		set of node properties, distinguishing attributes decorating nodes (\code{spec:att}), 
		attack detection rates decorating
		 attack nodes (functions $\mathcal{N}_a \rightarrow [0,1]$, \code{spec:dr}) and 
		defense effectiveness decorating defense nodes (functions $(\mathcal{N}_d \cup \mathcal{N}_c) \times \mathcal{N}_a \times \mathcal{A} \rightarrow [0,1]$, \code{spec:de}).
	\end{itemize}
%\end{definition}

We introduce the notion of \conf{} for a \ourtool{} model and equip it with an operational semantics based on rated transition systems. 
%
%\begin{definition}[\Conf]\label{sem:model}
	A \textit{\conf{}} of a \ourtool model $\mathcal{S}$ is a tuple $\langle C,s\rangle$, where $s$ is a state of attack behavior of $\mathcal{S}$ and $C$ is a set of constraints consisting of: 
	\begin{itemize}
\item all constraints of the model $\mathcal{S}$;
\item a predicate $\textit{has}(n)$ for each currently active node $n\in\mathcal{N}$;
\item constraints of form $\when{n_a} < \when{n_a'}$, for $n_a, n_a' \in \mathcal{N}_a$, denoting that $n_a$ was activated before $n'_a$, necessary to support \lil{OAND} refinements;  
\item an assignment of form $\textit{att}(n) = x$ for each attribute $\textit{att}$ and node $n \in \mathcal{N}$ to denote the value of the attribute for the node $n$, with $x \in \mathbb{R}$;  
\item assignments of form $\textit{value}_a(\textit{att}) = x$ and $\textit{value}_{\textit{def}}(\textit{att}) = x$ for each attribute $\textit{att}$ to denote its cumulative attacker and defender value, with $x \in \mathbb{R}$;%			\item For each attribute $\textit{att} \in \mathcal{AT}$, an assignment $\textit{att}_a = v$ and $\textit{att}_{\textit{def}} = v'$ for representing the attributes values for the attacker and the defender; 
\item an assignment of form $v = x$ for each variable $v \in \mathcal{V}$, with $x \in \mathbb{R}$; 
\item an assignment of form $\dr{n_a} = x$ for each attack node $n_a$ to denote detection rate of $n_a$, with $x \in \mathbb{R}$;  
\item a set $\textit{detect}(n_c)\subseteq \mathcal{N}_a$ for each countermeasure node $n_c$ to denote the attack nodes that can be detected by $n_c$. 
			\end{itemize}
%\end{definition}

Let $\mathcal{M}$ denote the set of all \conf{}s for a \ourtool model $\mathcal{S}$. We restrict to \conf{}s $\langle C,s \rangle$ such that $C$ is consistent, i.e.\ all constraints are satisfied, denoted by $\consistent{C}$. As we will see in Proposition~\ref{consistency-preservation}, this property is preserved by the operational semantics: no inconsistent \conf{} can be reached from a consistent one.  
We will use $\oplus$~to denote union of constraint sets, $\ominus$~for subtraction and $\vdash$~for entailment. 

\subsection{\ourtool{} Dynamics}
The dynamics of \ourtool \conf{}s is given as rated transition systems that specify how a \conf{} $\langle C,s \rangle$ can evolve into a \conf{} $\langle C',s' \rangle$ with a certain rate~$r$. Such evolution occurs as the consequence of the attacker trying to perform an action and the defender eventually reacting to mitigate it. We denote such an evolution with a transition of the form 
$
\langle C,s \rangle  \xrightarrow[]{r} \langle C',s' \rangle
$. 
In general, the dynamics is defined by a multi-relation $\rightarrow \subseteq \mathbb{N}^{\mathcal{M}\times \mathbb{R^+} \times \mathcal{M}}$ induced by the rules of Fig.~\ref{fig:sos}. We use a multi-relation since we have to account for multiple copies of the same transition with the same rate, as the probabilistic interpretation requires to \lq sum\rq\ such rates. 
Indeed, as we shall see, the dynamics of a configuration is ultimately defined as a discrete-time Markov chain, upon which the analysis of \ourtool is based.

\begin{figure*}%[t]
%\tiny
%\scriptsize
%\foonotetsize
\small
%\scalebox{0.9}{
	\if false
	\makeatletter
	\let\zz\Gscale@box
	\long\def\Gscale@box#1{%
		\xdef\thelastscalefactor{#1}%
		\zz{#1}}
	
	\noindent
	\resizebox{\textwidth}{!}{}
	\fi
	
		\noindent
	%\scalebox{0.77545}{
	\[
	\inference[\scriptsize\textsc{[Act]}]
	{
		s \xrightarrow{\textit{act\/},r,u,g} s' \qquad \textit{exe\/}(C,act,g) \qquad C'=u(C,\textit{act\/}) \qquad \consistent{C'} 
	}
	{
		\langle C,s \rangle \xrightarrow[]{r} \langle C', s' \rangle
	}
	\]
	%}
	\newline
	
	%\noindent
	%%\scalebox{0.77545}{
	%\[
	%\inference[\scriptsize\textsc{[Qry]}]
	%{
	%	s \xrightarrow{\textit{query\/}(n),r,u,g} s' \qquad \textit{exe\/}(C,query(n),g) \qquad \textit{has\/}(n) \in C \qquad C'=u(C,\textit{query\/}(n)) \qquad \consistent{C'} 
	%}
	%{
	%	\langle C,s \rangle \xrightarrow[]{r} \langle C', s' \rangle
	%}
	%\]
	%%}
	%\newline
	
		\noindent
	%\scalebox{0.77545}{
%	\[
%	\inference[\textsc{[Add]}]
%	{
%		s \xrightarrow{\textit{add}(n_a),u,g,r} s' \qquad exe(C,\textit{add}(n_a),g) \qquad K \subseteq \countermeasures{n_a,C} \\ C' = u(C,\textit{add}(n_a)) \oplus  \bigoplus_{n_c \in K} has(n_c) \oplus  has(n_a) \oplus  \bigoplus_{\{\,n \,\mid\, has(n) \in C\,\}} \when{n} < \when{n_a} \ominus\\ (\bigoplus_{\{\,n \,\mid\, n~\texttt{-RC->}~n_a\,\}} has(n)) \qquad
%	\qquad \consistent{C'}  
%	}
%	{
%		\langle C,s \rangle \xrightarrow[]{r \cdot \de{C'}{n_a,s} \cdot \prod\limits_{n_c \in K} \dr{n_c}{n_a} \cdot \prod\limits_{n_c \in (\countermeasures{n_a,C} \setminus K)} (1-\dr{n_c}{n_a}) } \langle C', s' \rangle
%	}
%	\]
	\[
	\inference[\scriptsize\textsc{[Add]}]
	{
		s \xrightarrow{\textit{add\/}(n_a),r,u,g} s' \qquad \textit{exe\/}(C,\textit{add\/}(n_a),g) \qquad\qquad \\ C' = u(C,\textit{add\/}(n_a)) \oplus \textit{has\/}(n_a) \oplus \bigoplus\limits_{n_c\,\in\,\countermeasures{n_a,C}} \hspace{-0.45cm}\textit{has\/}(n_c) \oplus     \bigoplus\limits_{\{\,n\,\in\,\mathcal{N}_a \,\mid\, \textit{has\/}(n)\,\in\,C\,\}} \hspace{-0.9cm}\when{n} < \when{n_a} \ominus \left(\bigoplus\limits_{\{\,n \,\mid\, n~\texttt{-RC->}~n_a\,\}} \hspace{-0.7cm}\textit{has\/}(n)\right)
\qquad 
\consistent{C'}  
	}
	{
		\langle C,s \rangle \xrightarrow[]{r \cdot \de{C'}{n_a,s} \cdot \dr{n_a}} \langle C', s' \rangle
	}
	\]
	%}
	\newline
	
	\noindent
	%\scalebox{0.77545}{
%	\[
%	\inference[\textsc{[Add]}]
%	{
%		s \xrightarrow{\textit{add}(n_a),u,g,r} s' \qquad exe(C,\textit{add}(n_a),g) \qquad K \subseteq \countermeasures{n_a,C} \\ C' = u(C,\textit{add}(n_a)) \oplus  \bigoplus_{n_c \in K} has(n_c) \oplus  has(n_a) \oplus  \bigoplus_{\{\,n \,\mid\, has(n) \in C\,\}} \when{n} < \when{n_a} \ominus\\ (\bigoplus_{\{\,n \,\mid\, n~\texttt{-RC->}~n_a\,\}} has(n)) \qquad
%	\qquad \consistent{C'}  
%	}
%	{
%		\langle C,s \rangle \xrightarrow[]{r \cdot \de{C'}{n_a,s} \cdot \prod\limits_{n_c \in K} \dr{n_c}{n_a} \cdot \prod\limits_{n_c \in (\countermeasures{n_a,C} \setminus K)} (1-\dr{n_c}{n_a}) } \langle C', s' \rangle
%	}
%	\]
	\[
	\inference[\scriptsize\textsc{[AddNoC]}]
	{
		s \xrightarrow{\textit{add\/}(n_a),r,u,g} s' \qquad \textit{exe\/}(C,\textit{add\/}(n_a),g) \qquad\qquad \\ C' = u(C,\textit{add\/}(n_a)) \oplus \textit{has\/}(n_a) \oplus \bigoplus\limits_{\{\,n\,\in\,\mathcal{N}_a \,\mid\, \textit{has\/}(n)\,\in\,C\,\}} \hspace{-0.9cm}\when{n} < \when{n_a} \ominus \left(\bigoplus\limits_{\{\,n \,\mid\, n~\texttt{-RC->}~n_a\,\}} \hspace{-0.75cm}\textit{has\/}(n)\right)
 \qquad	 
\consistent{C'}  
	}
	{
		\langle C,s \rangle \xrightarrow[]{r \cdot \de{C'}{n_a,s} \cdot (1-\dr{n_a})} \langle C', s' \rangle
	}
	\]
	%}
	\newline

		\noindent
	\[
	\inference[\scriptsize\textsc{[Fail]}]
	{
		s \xrightarrow{\textit{fail\/}(n_a),r,u,g} s' \qquad \textit{exe\/}(C,\textit{fail\/}(n_a),g) \qquad
C' = u(C,\textit{fail\/}(n_a)) \oplus  \bigoplus\limits_{n_c \in \countermeasures{n_a,C}} \hspace{-0.4cm}\textit{has\/}(n_c) 
\qquad 
\consistent{C'}
	}
	{
		\langle C,s \rangle \xrightarrow[]{r \cdot \de{C'}{n_a,s} \cdot \dr{n_a}} \langle C', s' \rangle
	}
	\]
	%}
	\newline
	
	\noindent
	%\scalebox{0.77545}{
	\[
	\inference[\scriptsize\textsc{[FailNoC]}]
	{
		s \xrightarrow{\textit{fail\/}(n_a),r,u,g} s' \qquad \textit{exe\/}(C,\textit{fail\/}(n_a),g) 
\qquad  C' = u(C,\textit{fail\/}(n_a)) \qquad \consistent{C'}  
	}
	{
		\langle C,s \rangle \xrightarrow[]{r \cdot \de{C'}{n_a,s} \cdot (1-\dr{n_a})} \langle C', s' \rangle
	}
	\]
	%}
	\newline

%		\noindent
%	%\scalebox{0.77545}{
%	\[
%	\inference[\textsc{[Fail]}]
%	{
%		s \xrightarrow{\textit{fail}(n_a),u,g,r} s' \qquad exe(C,\textit{fail}(n_a),g) \qquad K \subseteq \countermeasures{n_a,C} \\ C' = u(C,\textit{fail}(n_a)) \oplus  has(n_a) \oplus  \bigoplus_{\{\,n \,\mid\, has(n) \in C\,\}} \when{n} < \when{n_a}) \qquad
%	\qquad \consistent{C'}  
%	}
%	{
%		\langle C,s \rangle \xrightarrow[]{r \cdot \prod\limits_{n_c \in K} \dr{n_c}{n_a} \cdot \prod\limits_{n_c \in (\countermeasures{n_a,C} \setminus K)} (1-\dr{n_c}{n_a}) } \langle C', s' \rangle
%	}
%	\]
%	%}
%	\newline	
	
			\noindent
	%\scalebox{0.77545}{
	\[
	\inference[\scriptsize\textsc{[Rem]}]
	{
		s \xrightarrow{\textit{remove\/}(n_a),r,u,g} s' \qquad \textit{exe\/}(C,\textit{remove\/}(n_a),g) \qquad\qquad \\ C' = u(C,\textit{remove\/}(n_a)) \ominus \left( \textit{has\/}(n_a) \oplus  \bigoplus\limits_{\{\,n \,\mid\, \textit{has\/}(n) \in C\,\}} \hspace{-0.6cm}\when{n} < \when{n_a}\right)  \qquad \consistent{C'}  
	}
	{
		\langle C,s \rangle \xrightarrow[]{r} \langle C', s' \rangle
	}
	\]
	%}
	%\newline

\iffalse 	
	\centering
	\noindent
	%\scalebox{0.77545}{
	\[
	exe(C,a,g) =
	\begin{cases}
	\textit{false\/}   & \text{if $C \nvdash g$} \\
	\textit{false\/}   & \text{if $C = C' \oplus (\textit{do\/}(a) \rightarrow C'')$ and $C' \nvdash C''$} \\
	\textit{false\/}   & \text{if $a = \textit{add\/}(n_a)$ and $C = C' \oplus \textit{has\/}(n_a)$}\\
	\textit{false\/}   & \text{if $a = \textit{fail\/}(n_a)$ and $C = C' \oplus \textit{has\/}(n_a)$}\\
	\textit{false\/}   & \text{if $a = \textit{remove\/}(n_a)$ and $C = C' \oplus \neg \textit{has\/}(n_a)$}\\
	\textit{true\/}    & \text{otherwise\/}
	\end{cases}
	\]
\fi
%}
	%\vspace*{-0.15cm}
	\caption{\label{fig:sos}Operational semantics}
	%\vspace*{0.15cm}
\end{figure*}

The rules share some premises and effects. 
First, all rules need an attack behavior transition of the form $s \xrightarrow{\alpha,r,u,g} s'$, with current state of the attacker~$s$, action~$\alpha$, rate~$r$ and memory update~$u$, such that the executability conditions of the transition guard~$g$ hold. This is imposed by \textit{exe\/}($C,\alpha,g$), defined as: 
	\[\footnotesize
	\textit{exe\/}(C,\alpha,g)\!=\!%
	\begin{cases}
	\textit{false}\!\!&\!\!\text{if $C \nvdash g$} \\
	\textit{false}\!\!&\!\!\text{if $C = C' \oplus  (\textit{do\/}(\alpha) \rightarrow C'')$ and $C' \nvdash C''$} \\
	\textit{false}\!\!&\!\!\text{if $\alpha\!=\!\textit{add\/}(n_a)$ and $\textit{has\/}(n_a)\!\in\!C$, with $n_a\!\in\!\mathcal{N}_a$}\\
	\textit{false}\!\!&\!\!\text{if $\alpha\!=\!\textit{fail\/}(n_a)$ and $\textit{has\/}(n_a)\!\in\!C$, with $n_a\!\in\!\mathcal{N}_a$}\\
	\textit{false}\!\!&\!\!\text{if $\alpha\!=\!\textit{remove\/}(n_a\mkern-1mu)$ and $\textit{has\/}(n_a\mkern-1mu)\!\not\in\!C$,\,with $n_a\!\in\!\mathcal{N}_a$}\\
	\textit{true}\!\!&\!\!\text{otherwise\/}
	\end{cases}
	\]

Second, all rules require the resulting store to be consistent. 
Further conditions vary from rule to rule, as we will explain. 
By applying a rule on a \conf{} $\langle C, s \rangle$ due to a local transition $s \xrightarrow{\alpha,r,u,g} s'$, we obtain a \conf{} $\langle C', s' \rangle$, where $C'$ is obtained by applying the effects $u$ on the variables in $C$ (denoted $u(C,\alpha)$) and by possibly (de)activating nodes. In addition, $u(C,\alpha)$ updates cumulative attack and defense attribute values, as explained in Section~\ref{section:dsl}.  The semantics of $u(C,\alpha)$ is as expected, and not presented for conciseness. 

We now describe each rule in detail.

%Rules \textsc{Act} and \textsc{Qry} are straightforward. 
Rule \textsc{Act} executes user-defined actions: 
node activations are not altered by this rule so its effects are limited to variables. 
%
%The same applies to \textsc{Qry}, but it is allowed only if node $n$ is active, used by the attacker to test the activation of a node $n$. 
%

Rule \textsc{Add} is triggered by actions $\textit{add}(n_a)$: 
with probability $\textit{dr}(n_a)$, it may activate the set $\countermeasures{n_a,C}$ of countermeasure nodes able to detect $n_a$ that are not already active or inhibited by an active attack node $n'_a$. The set $\countermeasures{n_a,C}$ is defined\linebreak as follows, where $\texttt{-RC->}$ denotes a role-changing relation: 
%
%$$\countermeasures{n_a,C} = \{\,n_c \mid has(n_c) \not\in C \wedge n_a~\texttt{-RC->}~n_c \wedge \neg \exists n'_a . (has(n'_a) \in C \wedge n_c~\texttt{-RC->}~n'_a)\,\}$$, 
%\coma{This is wrong: I should not take the n_c using 'n_a~ \texttt{-RC->}', instead, I should take those countermeasures who might detect n_a}
%
\begin{multline*}%\[ 
%\countermeasures{n_a,C}=
\{\,n_c \mid n_a \in \textit{detect}(n_c) \wedge \textit{has}(n_c) \not\in C \wedge \neg \exists n'_a .\\ (\textit{has}(n'_a) \in C \wedge (n_c~\texttt{-RC->}~n'_a) \in C)\,\}
\end{multline*}%\]
%
%\todos{\coma{I can easily guess that $\lil{-RC->}$ is a role changing relation. But we did not define it. I informally defined it here, we can't define it everywhere because the tool does not support it}}
% In words, these are all countermeasures $n_c$ that can detect $n_a$ and that are not inhibited by an active attack node $n'_a$. 
%
Upon the execution of the rule, the constraint store is updated with the new attack node $n_a$, which is recorded to be the last active attack node of the store ($\when{n} < \when{n_a}$). Furthermore, the constraint store is also updated with each countermeasure node in $\countermeasures{n_a,C}$.
Another effect is that all defenses that have $n_a$ as opponent are deactivated. 
The rate of the obtained transition is not necessarily the original rate $r$ of the attack behavior transition. In fact, $r$ might be scaled by the defense effectiveness of the active defenses against $n_a$ in the newly obtained store, denoted by $\de{C'}{n_a,s}\in[0,1]$.   
We distinguish three cases: (i)~if $n_a$ has no role-changing relation, it is $1$; (ii)~if $n_a$ has a defense opponent $n_d$, it is the effectiveness of $n_d$ for $n_a$ and the current attacker; (iii)~if $n_a$ has a countermeasure opponent $n_c$, it is the product of the effectiveness of $n_c$ and that of any defense node that refines it, for $n_a$ and the attacker $\mathcal{A}$. 
Finally, we have to multiply the rate by the probability of activating the countermeasures, $\textit{dr}(n_a)$. 
Rule \textsc{AddNoC} is similar, but it covers the case in which the countermeasures $\countermeasures{n_a,C}$ do not get activated.

Rules \textsc{Fail} and \textsc{FailNoC} are similar to \textsc{Add} and \textsc{AddNoC}, but the attack node is not activated because they regard the \lil{fail} action which model failed attack attempts. 
Finally, rule \textsc{Rem} models the deactivation of an attack node. %The node is removed with the specified rate $r$ if all premises hold. 
%\comment{Cosi' come e' ora, gli attacchi ammazzano le difese. Se togli l'attacco, allora la difesa non torna a vita. Un'alternativa sarebbe che gli attacchi inibiscono le difese: se gli attacchi vanno via le difese tornano attive.}

It is easy to see that the semantic rules ensure consistency is preserved along sequences of configurations, since consistency is a premise in every rule, and hence in every transition.

\newtheorem{proposition}{Proposition}

\begin{proposition}
\label{consistency-preservation}
Let $\mathcal{S}$ be a \ourtool model and $\langle C , s \rangle$ be a configuration such that $C$ is consistent. Then for any configuration $\langle C' , s' \rangle$ such that $ \langle C , s \rangle \rightarrow^* \langle C' , s' \rangle $ it holds that $C'$ is consistent.
\end{proposition}

The probabilistic interpretation of rated transition systems yields DMTCs. 
A DTMC is a tuple $\langle \Gamma, \Pi \rangle$ where $\Gamma$ is a set of states and 
$\Pi : \Gamma \rightarrow [0,1]$ is a probability transition function, i.e.\ such that for all $s \in \Gamma$, $\sum_{s' \in \Gamma}\, \Pi(s,s') = 1$.
The DTMC semantics of a rated transition system is obtained by normalising the rates into $[0..1]$ such that in each state/configuration, the sum of the rates of its outgoing transitions equals one. So, for a rated transition system $\rightarrow$ on a set of configurations $\mathcal{M}$ we obtain the DTMC $\langle \mathcal{M} , \Pi \rangle$ where, for each pair of states $s,s' \in \mathcal{M}$, the probability transition function $\Pi$ is defined by 
%
%\[ 
%\Pi(s,s') =  \frac{\sum_{(s,r,s') \in \rightarrow}\,r}{\sum_{(s,r,s'') \in \rightarrow,\,s'' \in \mathcal{M}}\,r}
%\]
\[ 
\Pi(s,s') =  
\begin{cases}
\frac{\sum_{(s,r,s') \in \rightarrow}\,r}{\mathbf{out}(s)} & \text{ if } \mathbf{out}(s)>0
\\
1 & \text{ if } \mathbf{out}(s)=0 \text{ and } s=s'
\\
0 & \text{ otherwise }
\end{cases}
\]
where $\mathbf{out}$ denotes the outdegree of a configuration. 
Note that self-loops with probability $1$ are added to configurations without outgoing transitions.
The DTMC semantics of \ourtool models is used in our analyses, described in the next section. 

\section{\ourtool{} Supported Quantitative Analyses}
\label{section:analysis}

\ourtool supports the quantitative analysis of probabilistic attack scenarios by means of statistical model checking (SMC)~\cite{Agha18,LLTYSG19} as well as probabilistic model checking (PMC)~\cite{BK08}, thus providing additional analysis capabilities to what other risk analysis tools typically offer. 

SMC is concerned with running a sufficient number of (probabilistic) simulations of a system model to obtain statistical evidence (with a predefined level of statistical confidence) for the quantitative properties to be checked. 
Compared to obtaining exact results (with 100\% confidence) with exact analysis techniques like (probabilistic) model checking, SMC offers unique advantages over exhaustive (probabilistic) model checking. Most importantly, SMC scales better. First, there is no need to generate entire state spaces, thus avoiding the combinatorial state-space explosion problem typical of model checking~\cite{CHVB18}. Second, the set of simulations to be carried out can be trivially  distributed and run in parallel, thus scaling better with hardware resources. \mv, indeed, can be run on multi-core machines, clusters or distributed computers with a nearly linear speedup. Another advantage concerns its uptake in industry. Compared to model checking, SMC is simple to implement, understand and use, and it requires no specific modeling effort other than a system model that can be simulated and checked against quantitative properties. In fact, SMC is more and more being applied in industry~\cite{DBLP:conf/ifm/GilmoreTV14,BLLV16,Filipovikj2016SimulinkTU,ABFMSLQE17,BDG17,CLSQTL17,BBC18,PFG18,BADFL19,BBFL19,DBLP:conf/rssrail/BasileFRM19,FMBBF20,GBP20}.

In \ourtool, the SMC analysis is obtained thanks to the internal DTMC simulator with \mv{}~\cite{SV13,GRV17}, a framework for enriching simulators with SMC capabilities, while the PMC analysis is obtained thanks to \ourtool's DTMC exporting capabilities in a format supported by PRISM~\cite{KNP11} and STORM~\cite{DJKV17}.
SMC is necessary because the \ourtool DSL has high expressivity, allowing for potentially unbounded variables and high variability in the models, thus often giving rise to large or infinite state spaces. 
PMC can instead be used for models with finite state spaces for exact analyses.

Next we showcase two SMC analysis capabilities of \ourtool on our running example. PMC cannot be used in this case as the model has an infinite state space. We will showcase PMC analyses using PRISM in Section~\ref{section:validation}. 

%\paragraph{Analysis while varying simulation steps}
\subsection{Analysis while Varying Simulation Steps}
We start by studying the probabilities of activating attacks and countermeasures while varying the simulation step. This is expressed in \code{spec:analist}: The pattern \lil{from}-\lil{to}-\lil{by} specifies that we are interested in the first $100$~steps. 
We list $8$~properties of interest ($1$ per attack node, considering \codefont{Find}\-\codefont{Code1}  active, plus countermeasure \codefont{Lock}\-\codefont{Down}). 
Each property can be an arithmetic expression of nodes (evaluating to~$1$ or~$0$ if the node is active or not, respectively), variables or attributes. The properties are considered in all $100$~steps, totaling $800$~properties. 
%
%\begin{wrapfigure}{r}{0.535\textwidth} 
%\vspace{-0.8cm}
\begin{lstlisting}[language=spec,caption={Analysis of the scenario},captionpos=b, label={spec:analist},escapeinside={@}{@},numbers=none]
begin analysis@\label{anafrom1}@
 query = eval from 1 to 100 by 1 : 
  {RobBank, OpenVault, BlowUp, LearnCombo, GetToVault, 
   FindCode2, FindCode3, LockDown}
 default delta = 0.1 alpha = 0.1 parallelism = 1
end analysis@\label{anafrom2}@
\end{lstlisting}
%\vspace{-1.1cm}
%\vspace{-0.5cm}\end{wrapfigure}
%

Each such actual property~$p_i$ denotes a random variable~$X_i$ which gets a real value assigned in each simulation. 
\mv{} estimates the expected value $E[X_i]$
 of each of the $800$~properties (reusing the same simulations) as the mean $\overline{x}_i$ of $n$ independent simulations, with $n$ large enough to guarantee an $(\alpha,\delta)$ \emph{confidence interval} (CI): $E[x_i]$ 
belongs to $[\overline{x}_i - \sfrac{\delta\!}{2}, \overline{x}_i + \sfrac{\delta\!}{2}]$ with statistical confidence $(1-\alpha)\cdot100\%$. 
%
% A CI is thus specified in terms of two parameters: $\alpha$ and $\delta$.
%
The CI is given by \lil{alpha} and \lil{default} \lil{delta} (but a property-specific $\delta$ could be used instead).
Finally, \lil{parallelism} states how many local processes should be launched to distribute the simulations.
Overall the analysis %of these 800 properties 
required $400$~simulations, performed in $16$~seconds on a standard laptop machine.

\begin{figure}[t]
\centering
\includegraphics[width=\linewidth]{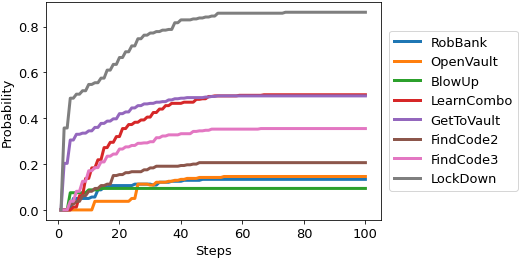}
\vspace*{-0.25cm}
\caption{\label{fig:analysis}Analysis result of the properties in \code{spec:analist}}
\end{figure}

Fig.~\ref{fig:analysis} shows the results. Recall (Fig.~\ref{fig:structure},  \code{spec:hcon}) that \codefont{Rob}\-\codefont{Bank} requires \codefont{Open}\-\codefont{Vault} or \codefont{Blow}\-\codefont{Up}. 
The probability to activate \codefont{Rob}\-\codefont{Bank} starts growing after step 4, stabilizing at $0.17$, while those of \codefont{Open}\-\codefont{Vault} and \codefont{Blow}\-\codefont{Up} reach~$0.15$ and $0.11$, resp. 
We know from \code{spec:acon} that they cannot both be activated, so one should be able to activate \codefont{Rob}\-\codefont{Bank} with probability almost~$0.26$.  
Instead, the actual probability is scaled down by $\frac{2}{3}$ due to the probabilistic choice from \codefont{start} to \codefont{complete} in Fig.~\ref{fig:behavior}: \codefont{Rob}\-\codefont{Bank} can either succeed or fail. 

Note that \codefont{Lock}\-\codefont{Down} has a high probability to be activated, reaching about~$0.85$ after $60$~steps. This is coherent with \code{spec:de}, stating that any \codefont{Blow}\-\codefont{Up} attempt is detected. 
One might expect the probability to activate \codefont{Blow}\-\codefont{Up} to be higher than that of \codefont{Lock}\-\codefont{Down}, as the former triggers the latter. However, this is not true. 
This is explained by the fact that both succeeded and failed \codefont{Blow}\-\codefont{Up} attempts are detected (cf.\ success \lil{succ}(\codefont{BlowUp}) and failure \lil{fail}(\codefont{BlowUp}) actions in Fig.~\ref{fig:behavior}). 
Interestingly, if we added \codefont{Laser}\-\codefont{Cutter} to the initial configuration, then the probability of activating \codefont{Lock}\-\codefont{Down} would remain~$0$, as it is inhibited by \codefont{Laser}\-\codefont{Cutter}. 

%\paragraph{Analysis at the verification of a condition}
\subsection{Analysis at the Verification of a Condition}
We can also compute properties evaluated as soon as a given condition verifies.
Here we %We exemplify this by computing 
compute the probability for each attack node to be the first attempted and succeeded, as well as the average number of steps needed to perform the first attempt.
\code{spec:analistwhen} expresses these $9$~properties ($1$ probability per attack node plus the average number of steps). 
Note that the \lil{from}-\lil{to}-\lil{by} pattern is replaced by %keyword 
\lil{when} to specify that the properties should be evaluated in the first state satisfying \codefont{AttackAttempts} \codefont{==} \codefont{1}. 
Moreover, the list of properties of interest now includes \lil{steps}, for which we give a specific delta, evaluated as the average number of steps computed to reach the first state satisfying the required condition. 

\begin{lstlisting}[language=spec,caption={Analysis of the scenario},captionpos=b, label={spec:analistwhen},escapeinside={@}{@},numbers=none]
begin analysis
 query = eval when {AttackAttempts == 1} : 
  {RobBank, OpenVault, BlowUp, LearnCombo, GetToVault, 
   FindCode2, FindCode3, LockDown, steps[delta = 0.5]}
 default delta = 0.1 alpha = 0.1 parallelism = 1
end analysis
\end{lstlisting}

\setlength{\tabcolsep}{4.25pt}
\begin{table}[hb]
\centering
%\vspace*{-0.15cm}
\scalebox{0.9}{
\begin{tabular}{cccccccccc}
\toprule
\texttt{Rob} & \texttt{Open} & \texttt{Blow} & \texttt{Learn} & \texttt{GetTo} & \texttt{Find} & \texttt{Find} & \texttt{Lock} & \multirow{2}{*}{\texttt{steps}} \\ 
\texttt{Bank} & \texttt{Vault} & \texttt{Up} & \texttt{Combo} & \texttt{Vault} & \texttt{Code2} & \texttt{Code3} & \texttt{Down} & \\ \midrule
%0 & 0 & 0 & 0 & 0.23 & 0 & 0.04 & 0.32 & 2.61\\
0 & 0 & 0 & 0 & 0.27 & 0 & 0.01 & 0.32 & 2.51\\ 
\bottomrule\\
\end{tabular}
}
\caption{\label{tab:runningwhen}Analysis result of the properties in \code{spec:analistwhen}} %\lil{when}}
\end{table}

Overall, the analysis required $400$~simulations, performed in a few seconds on a standard laptop machine.
The analysis results are provided in Table~\ref{tab:runningwhen}. 
The first four attack nodes have probability $0$ of being the first attempted and succeeded attack. This is coherent with the diagram in Fig.~\ref{fig:structure}, as such attacks are not leaves of the diagram and thus require other attacks to succeed first. 
Consistently with Fig.~\ref{fig:analysis}, \codefont{GetToVault} has higher probability than \codefont{FindCode2} and \codefont{FindCode3}. 
Intuitively, this depends on the way the attacker's behavior is defined. As sketched in Fig.~\ref{fig:behavior} and specified in \code{spec:attackers:behavior}, starting from state \codefont{start} we only have to perform one step to try \codefont{GetToVault} attacks, while to try finding a code requires traversing two more states, in each of which other competing actions are enabled.
In turn, \codefont{FindCode2} has lower probability (belonging to the interval $[0,0.05]$ due to the imposed CI) than \codefont{FindCode3} due to the defense \mbox{\codefont{Memo}}. 
Interestingly, we note a probability of~$0.32$ of activating the countermeasure \codefont{LockDown}. This means that failed \codefont{BlowUp} attempts were detected.
Finally, Table~\ref{tab:runningwhen} also shows that, on average, $2.51$~steps are needed to perform one attack attempt. Indeed, in state \codefont{start} no attack attempt is allowed, so two steps are needed to attempt  \codefont{GetToVault} or \codefont{BlowUp} attacks, while three are needed for \codefont{FindCode} attempts.

%\paragraph{Simulating and exporting} 
\subsection{Simulating and Exporting} 

\ourtool models can be debugged by performing probabilistic simulations.  
%
%\begin{wrapfigure}{r}{0.41\linewidth}
%%\vspace*{-0.1cm}
%\begin{lstlisting}[language=spec,caption={Log generation},captionpos=b, label={spec:debug},escapeinside={@}{@},numbers=none]
%begin simulate
% seed = 1 steps = 1
% file = "simulation.log"
%end simulate
%\end{lstlisting}
%%\vspace{-0.25cm}
%\end{wrapfigure}
%%
% 
\code{spec:debug} prints (in file \codefont{sim.log}) all chosen states and other useful information of the simulation suitable for debugging. 
\ourtool's \codefont{DTMC Exporter} can generate  entire %explicit 
DTMCs and export them in the input format accepted by the probabilistic model checkers PRISM or STORM. 

%\begin{wrapfigure}{r}{0.41\linewidth}
%%\vspace{-0.2cm}
%\begin{lstlisting}[language=spec,caption={DTMC export},captionpos=b, label={spec:DTMC},escapeinside={@}{@},numbers=none]
%begin exportDTMC
% file = "RobBank.pm"
% label with "hasRB"
%       when has(RobBank)
%end exportDTMC
%\end{lstlisting}
%\vspace{-0.25cm}
%\end{wrapfigure}
%
\code{spec:DTMC} %instead 
shows how to export the DTMC of our running example for external analysis, labeling with \codefont{"hasRB"} all states in which a \codefont{RobBank} attack succeeded. 

\noindent\begin{minipage}{\columnwidth}
%\centering
\begin{minipage}[t]{0.32\linewidth}
%\begin{wrapfigure}{r}{0.41\linewidth}
%\vspace*{-0.1cm}
\begin{lstlisting}[language=spec,caption={Log\,generation},captionpos=b, label={spec:debug},escapeinside={@}{@},numbers=none]
begin simulate
 seed = 1 steps = 1
 file = "sim.log"
end simulate
\end{lstlisting}
%\vspace{-0.25cm}
%\end{wrapfigure}
%
\end{minipage}
\quad
%\centering
\begin{minipage}[t]{0.62\linewidth}
%\begin{wrapfigure}{r}{0.41\linewidth}
%\vspace{-0.2cm}
\begin{lstlisting}[language=spec,caption={DTMC export},captionpos=b, label={spec:DTMC},escapeinside={@}{@},numbers=none]
begin exportDTMC
 file = "RobBank.pm"
 label with "hasRB" when has(RobBank)
end exportDTMC
\end{lstlisting}
%\vspace{-0.25cm}
%\end{wrapfigure}
\end{minipage}
\end{minipage}

\section{Validation of \ourtool{}}
\label{section:validation}

A variety of extensions of attack-tree models exist and no single approach has so far emerged as the ultimate solution~\cite{whitepaper,survey,HKCH17,WAFP19}. 
This section shows the flexibility of \ourtool by illustrating how features from three seminal and influential kinds of attack trees can be specified in \ourtool, and how the latter's analysis capabilities can be used to complement and enrich the analyses provided by existing tools.
All \mv{} analyses in this section used 0.1 for both $\alpha$ and $\delta$.
The tool, its source code and the models and analyses are available at {\color{blue}\url{https://github.com/risqflan/RisQFLan/wiki}}.

\subsection{Case Study 1: Ordered Attacks}
\label{section:case1}
This section shows that the \ourtool DSL can be used to model features from \emph{enhanced attack trees}, an extension of basic attack trees proposed in~\cite{CY07}, and that \ourtool hence complements the analysis capabilities of~\cite{CY07} with (exact) PMC and SMC %statistical model checking 
on specific attacker profiles.
We do so by illustrating how \emph{ordered attacks}, a key differentiating feature of such \emph{enhanced attack trees\/}, can be specified in \ourtool.

\subsubsection{Ordered Attacks to \lq\lq Bypassing 802.1x\rq\rq}

As illustrative example, we use one case study from~\cite{CY07}, %modeled and evaluated in~\cite{CY07}, 
namely an enhanced attack tree modeling complex (ordered) attacks on wireless LANs using protocol IEEE 802.11. 
Fig.~\ref{figure:Bypassing802originalAttackTree}, reproduced from~\cite{CY07}, illustrates the enhanced attack tree. The main idea is that the authentication mechanism of the protocol can be compromised through hijacking authenticated sessions~(\codefont{B}) or man-in-the-middle attacks~(\codefont{E}). The sub-trees of~\codefont{B} and~\codefont{E} further refine both attacks into specific sub-goals. % and, ultimately, attacker actions.
%This can be represented in \ourtool as shown in encoded in \code{code:Bypassing802}.

\noindent\begin{minipage}{\columnwidth}
%\centering
\begin{minipage}[t]{0.47\linewidth}
\begin{figure}[H]
%\begin{wrapfigure}{r}{0.50\linewidth}
\hspace*{1cm}\includegraphics[width=.75\linewidth]{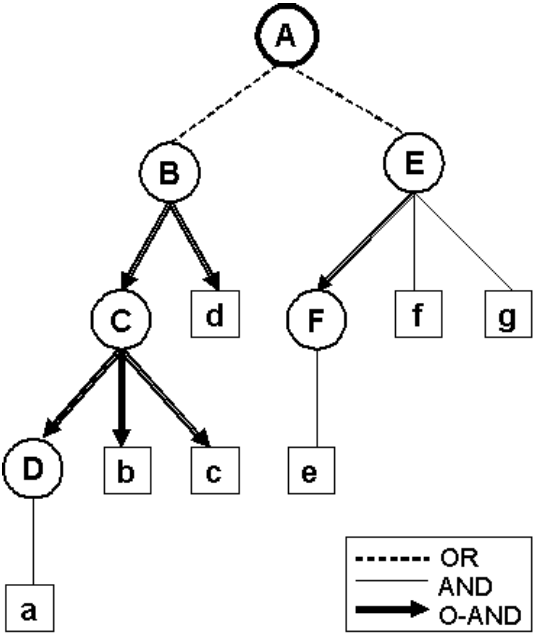}
\vspace{-0.5cm}
\caption{\label{figure:Bypassing802originalAttackTree}Enhanced attack tree for \lq\lq Bypassing 802.1x\rq\rq~\cite{CY07}}
%\vspace{-0.25cm}\end{wrapfigure}
\end{figure}
\end{minipage}
\qquad\quad\,\
\begin{minipage}[t]{0.35\linewidth}
%\begin{wrapfigure}{r}{0.4\linewidth}
%\ \\[-.2em]
\vspace{1cm}\begin{lstlisting}[language=spec,caption={Attack tree of Fig.~\ref{figure:Bypassing802originalAttackTree} in \ourtool},captionpos=b,abovecaptionskip=20pt,
label={code:Bypassing802},numbers=left,escapeinside={@}{@}]
begin attack diagram
 A -OR-> {B, E} 
 B -OAND-> [C, d] @\label{bypass1}@
 C -OAND-> [D, b, c] @\label{bypass2}@
 D -AND-> {a}
 E -OAND-> [F, fg] @\label{bypass3}@
 fg -AND-> {f, g} @\label{bypass4}@
 F -> {e}
end attack diagram
\end{lstlisting}
\vspace{0.5cm}%\end{wrapfigure}
\end{minipage}
\end{minipage}

\subsubsection{Specifying Ordered Attacks in \ourtool{}}

\code{code:Bypassing802} shows a model of the enhanced attack tree of Fig.~\ref{figure:Bypassing802originalAttackTree} in \ourtool. It is worth observing how the ordering relation is modeled. %In particular, 
The original model in Fig.~\ref{figure:Bypassing802originalAttackTree} prescribes that: 
%\begin{itemize}
%
%\item
(i)~to achieve attack~\codefont{B}, sub-goal~\codefont{C} must be achieved before~\codefont{d} (cf.\ \lline{bypass1} in \code{code:Bypassing802});
%
%\item
(ii)~to achieve attack \codefont{C}, sub-goal \codefont{D} must be achieved before \codefont{b}, which itself must be achieved before~\codefont{c} (cf.\ \lline{bypass2} in \code{code:Bypassing802});
%
%\item
and (iii)~to achieve attack~\codefont{E}, sub-goal~\codefont{F} must be achieved before~\codefont{f} and~\codefont{g} (cf.\ \llines{bypass3}{bypass4} in \code{code:Bypassing802}). Note that in the \ourtool specification, auxiliary node~\codefont{fg} is used to group the unordered conjunction of %nodes 
\codefont{f} and~\codefont{g}. 
%\end{itemize}
%

\subsubsection{Complementing the Analysis of~\cite{CY07} with \ourtool{}}

The main analysis feature of the approach in~\cite{CY07} consists of inspecting activity logs to recognise potential attacks as per the specified enhanced attack trees. 
With \ourtool this can be augmented with exact or statistical probabilistic verification on the average behaviour of specific attacker profiles. To illustrate this we modeled four attacker profiles:
\begin{description}
\item{\codefont{Best}:} an attacker that knows one of the optimal order of attacks to perform to achieve the main attack goal;
\item{\codefont{AverageA}:} an attacker %that 
randomly trying %tries 
attacks until achieving the main attack goal %is achieved 
or a wrong order %has 
led to failure;
\item{\codefont{AverageB}:}$\mkern2mu$like\,\codefont{AverageA}\,but\,can\,undo\,attacks\,(backtrack); 
\item{\codefont{Worst}:} like \codefont{AverageA} but chooses attacks with a probability inversely proportional to the order used by \codefont{Best}.
\end{description}

\begin{figure}[ht]
\centering\includegraphics[width=.8\columnwidth]{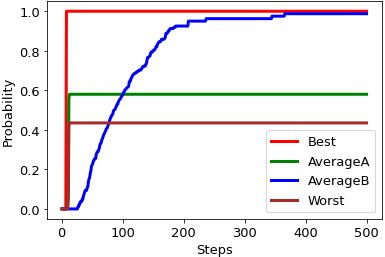}
%\includegraphics[width=0.45\columnwidth]{figures/Bypassing802plot-Best}
%\includegraphics[width=0.45\columnwidth]{figures/Bypassing802plot-Worst}\\
%\includegraphics[width=0.45\columnwidth]{figures/Bypassing802plot-AverageA}
%\includegraphics[width=0.45\columnwidth]{figures/Bypassing802plot-AverageB}
%\vspace*{-0.25cm}
\caption{Statistical analysis on \lq\lq Bypassing 802.1x\rq\rq}
\label{figure:plot802}
\end{figure}

Fig.~\ref{figure:plot802} presents the results of SMC analysis of each such attacker profile, showing that they converge to different attack success probabilities. %The results have been obtained with SMC. 
We have also exported the corresponding DTMCs and analysed them with PMC using PRISM. 
PRISM computed the same results for all attackers except for \codefont{AverageB}, whose DTMC is too large  (due to backtracking in the attacker's strategy) for PRISM or STORM to be able to handle it. 
%When PRISM was able to handle the DTMCs, the obtained results confirmed those of \mv. In one case (\codefont{AverageB}) the DTMC is too large (due to backtracking in the attacker's strategy) for PRISM to be able to handle it. 
%
Attackers \codefont{Best} and \codefont{AverageB} obviously achieve the attack with probability~1, although the latter needs more time. The \codefont{AverageA} attacker is next, achieving a success probability slightly above~$0.6$, while the \codefont{Worst} attacker achieves an attack with probability about~$0.4$.

\subsection{Case Study 2: Noticeability}
\label{section:case2}

This section shows that the \ourtool DSL can be used to 
model features from \emph{capabilities-based attack trees}~\cite{whitepaper}, an extension of basic attack trees offered in the commercial attack tree-based risk assessment tool Secur\emph{IT}ree~\cite{securitree}. This means \ourtool complements the models of Secur\emph{IT}ree with explicit dynamic attack behavior\,\footnote{Amenaza has similar plans for Secur\emph{IT}ree v5.1 (T. Ingoldsby, personal communication, April 1, 2020).} and its analysis capabilities with analysis of attacker profiles. We illustrate how the notion of \emph{noticeability}, one of the capability features of \emph{capabilities-based attack trees}, can be specified in \ourtool.
 
\subsubsection{Noticeability Capabilities of BurgleHouse}

\begin{figure}[ht]
\begin{minipage}[c]{0.33\columnwidth}
\includegraphics[width=.95\linewidth]{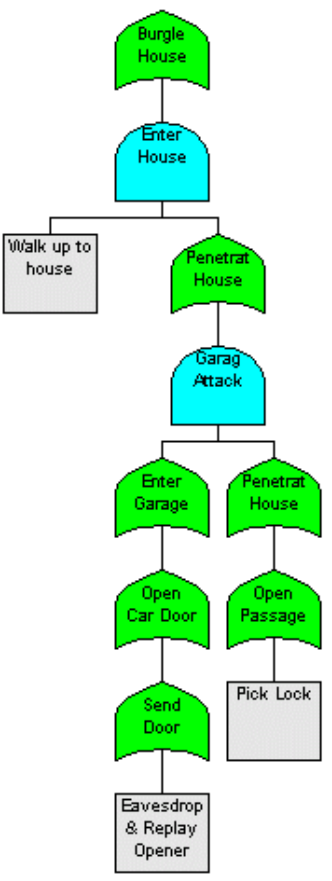}
\end{minipage}
\begin{minipage}[b][3.75cm][t]{0.105\columnwidth}
\includegraphics[width=.95\linewidth]{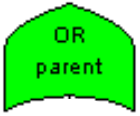}
\\[.6em] 
\includegraphics[width=.95\linewidth]{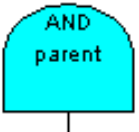}
\end{minipage}
\begin{minipage}[c]{0.53\columnwidth}
\includegraphics[width=.95\linewidth]{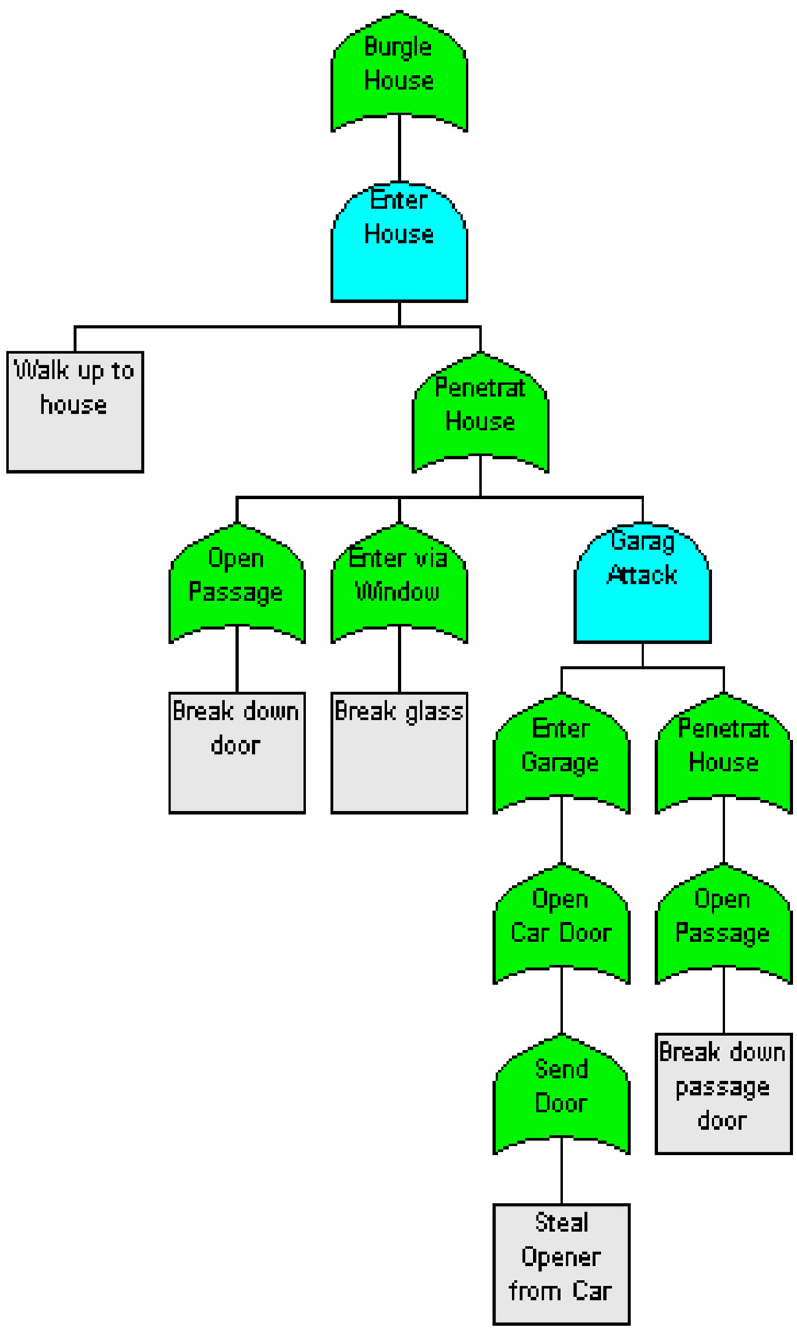}
\end{minipage}
\caption{Capabilities-based attack trees: ``Cat Burglar'' (left) and ``Juvenile Delinquent'' (right)~\cite{securitree}}
\label{figure:BurgleHouseoriginalAttackTrees}
\end{figure}

As illustrative example, we use two attack scenarios studied %modeled and evaluated 
in~\cite{securitree}, namely the Cat Burglar and Juvenile Delinquent scenarios from the BurgleHouse case study. Fig.~\ref{figure:BurgleHouseoriginalAttackTrees}, reproduced from~\cite{securitree}, depicts two capabilities-based attack trees which can be easily encoded in the RisQFlan DSL using \lil{OR} and \lil{AND} refinements. The idea is that a house can be burglarized by entering the house by carrying out two sub-goals: \codefont{WalkUpToHouse} and \codefont{PenetrateHouse}. The latter is further refined into sub-goals.  
In the Cat Burglar scenario the house can only be penetrated via a \codefont{GarageAttack}, whereas in the Juvenile Delinquent scenario there are two further alternatives: opening the passage door by breaking it down or entering via the window by breaking the glass.
We consider one of the three so-called behavioral indicators associated to attacker actions in~\cite{securitree}, namely \emph{noticeability}. The values were kindly provided by Terry Ingoldsby of Amenaza Technologies Ltd.\ together with a license for Secur\emph{IT}ree v5.0.
%through passage doors, windows, garage, walls (incl.\ the roof), chimney, floor, or social engineering (convincing the resident). 

%\begin{figure}[ht]
%\includegraphics[width=0.5\columnwidth]{figures/CatBurglaroriginalAttackTree}
%\caption{Capabilities-based attack tree for \lq\lq Cat Burglar\rq\rq~\cite{securitree}}
%\label{figure:JuvenileDelinquentorigCatBurglaroriginalAttackTreeinalAttackTree}
%\end{figure}

%\begin{figure}[ht]
%\includegraphics[width=0.5\columnwidth]{figures/JuvenileDelinquentoriginalAttackTree}
%\caption{Capabilities-based attack tree for \lq\lq Juvenile Delinquent\rq\rq~\cite{securitree}}
%\label{figure:JuvenileDelinquentoriginalAttackTree}
%\end{figure}

\subsubsection{%Modeling 
Noticeability in \ourtool{}}

%Code~\ref{code:CatBurglar} shows how the attack detection rates of the Cat Burglar scenario are modeled in \ourtool. 
%
Codes~\ref{code:CatBurglar} and~\ref{code:JuvenileDelinquent} show how the noticeability values of the Cat Burglar and Juvenile Delinquent scenarios, resp., are modeled as a \codefont{Noticeability} attribute in \ourtool: 
%Not surprisingly, 
walking up to the house is almost unnoticeable, %(\lline{unnote1} in Codes~\ref{code:CatBurglar}, \ref{code:JuvenileDelinquent}), 
while breaking a door or glass is more noticeable. %(\lline{note} in Code~\ref{code:JuvenileDelinquent}).

\begin{lstlisting}[language=spec,caption={Noticeability of \lq\lq Cat Burglar\rq\rq\ specified in \ourtool},captionpos=b, label={code:CatBurglar},numbers=left,escapeinside={@}{@}]
begin attributes
 Noticeability = {WalkUpToHouse = 0.01, @\label{unnote1}@
  EavesdropAndReplayOpenerCode = 0.05, PickLock = 0.02}
end attributes
\end{lstlisting}
%\vspace*{-0.25cm}%
\begin{lstlisting}[language=spec,caption={Noticeability of \lq\lq Juvenile Delinquent\rq\rq\ %specified 
in \ourtool},captionpos=b, label={code:JuvenileDelinquent},numbers=left,escapeinside={@}{@}]
begin attributes
 Noticeability = {WalkUpToHouse = 0.01, @\label{unnote2}@
  BreakDownDoor = 0.3, BreakGlass = 0.3, @\label{note}@
   StealOpenerFromCar = 0.2, BreakDownPassageDoor = 0.1}
end attributes
\end{lstlisting}

\subsubsection{Complementing the Analysis of~\cite{securitree} with \ourtool{}}

One of the analysis features of Secur\emph{IT}ree consists of the possibility to identify attack scenarios according to one or more behavioral indicators. For instance, by pruning the complete attack tree of the BurgleHouse case study with $29$~nodes, Secur\emph{IT}ree identified the above scenarios as corresponding to the specific capabilities of threat agents of the Cat Burglar and Juvenile Delinquent type (which avoid attacks that involve a risk of getting caught greater than 10\% and 30\%, resp., expressed through the noticeability criterion). Similarly, \ourtool can limit its analysis to such type of scenarios by imposing quantitative constraints (cf.\ Code~\ref{spec:qcon} in Section~\ref{section:dsl}).
%
%\begin{lstlisting}[language=spec,caption={Quantitative constraint for \lq\lq Juvenile Delinquent\rq\rq},captionpos=b, label={spec:qconbis},numbers=none]
%begin quantitative constraints
% { value(Cost) <= 0.3 }
%end quantitative constraints
%\end{lstlisting}
%
However, \ourtool can also augment such analyses with quantitative verification on the average behavior of specific attacker profiles as well as with estimation of the average noticeability of specific (successful) attacks. To illustrate this, we modeled four attacker profiles:
\begin{description}
\item{\codefont{Best}:} an attacker that knows an optimal, most unnoticeable order$\mkern3.5mu$of$\mkern3.5mu$attacks$\mkern3.5mu$to$\mkern3.5mu$perform$\mkern3.5mu$to$\mkern3.5mu$achieve$\mkern3.5mu$the$\mkern3.5mu$main$\mkern3.5mu$attack$\mkern3.5mu$goal;
\item{\codefont{AverageA}:} an attacker that randomly tries attacks until the main attack goal is achieved;
\item{\codefont{AverageB}:}$\mkern2mu$like\,\codefont{AverageA}\,but\,can\,undo\,attacks\,(backtrack); 
\item{\codefont{Worst}:} like \codefont{AverageA} but chooses attacks with a probability inversely proportional to \codefont{Best}.
\end{description}

\begin{figure}[ht]
 \centering
 \includegraphics[width=0.8\columnwidth]{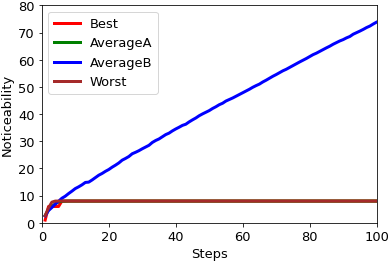}
 %\includegraphics[width=0.6\columnwidth]{figures/Placeholder}
 %\includegraphics[width=0.45\columnwidth]{figures/CatBurglarplot-Best+valuex100}
 %\includegraphics[width=0.45\columnwidth]{figures/CatBurglarplot-Worst+valuex100}\\
 %\includegraphics[width=0.45\columnwidth]{figures/CatBurglarplot-AverageA+valuex100}
 %\includegraphics[width=0.45\columnwidth]{figures/CatBurglarplot-AverageB+valuex100}
 %\vspace{-0.25cm}
 \caption{Statistical analysis on \lq\lq Cat Burglar\rq\rq}
 \label{figure:plotCatBurglar}
 \end{figure}

\begin{figure}[ht]
 \centering
 \includegraphics[width=0.8\columnwidth]{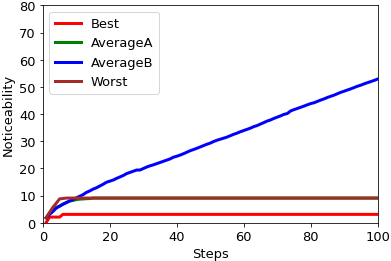}
 %\includegraphics[width=0.75\columnwidth]{figures/JuvenileDelinquentplot}
 %\includegraphics[width=0.6\columnwidth]{figures/Placeholder}
 %\includegraphics[width=0.45\columnwidth]{figures/JuvenileDelinquentplot-Best+valuex10}
 %\includegraphics[width=0.45\columnwidth]{figures/JuvenileDelinquentplot-Worst+valuex10}\\
 %\includegraphics[width=0.45\columnwidth]{figures/JuvenileDelinquentplot-AverageA+valuex10}
 %\includegraphics[width=0.45\columnwidth]{figures/JuvenileDelinquentplot-AverageB+valuex10}
 %\vspace{-0.25cm}
 \caption{Statistical analysis on \lq\lq Juvenile Delinquent\rq\rq}
 \label{figure:plotJuvenileDelinquent}
 \end{figure}

We analysed these 4 attackers in the two scenarios using the SMC analysis capabilities of \ourtool. 
%Fig.~\ref{figure:plotCatBurglarJuvenile}
Fig.~\ref{figure:plotCatBurglar} and Fig.~\ref{figure:plotJuvenileDelinquent} 
show how the attacker profiles converge to different average noticeability values of the attacks\,\footnote{To make the differences visible, the noticeability values of the Cat Burglar and Juvenile Delinquent scenarios were multiplied by~$10$ and~$100$, resp.}. 
Note that, contrary to the case study presented in the previous section, none of the orders of attacks can result in failure. In fact, while not shown, in both scenarios all attackers succeed with probability~$1$, although in both cases attacker \codefont{AverageB} needs considerably more time.
Moreover, in the Cat Burglar scenario, all successful attackers that cannot backtrack use the same set of actions. In fact, the average noticeability value of the \codefont{Best}, \codefont{AverageA}, and \codefont{Worst} attackers is~$8$, whereas the repeated attack attempts of the \codefont{AverageB} attacker guarantee that (s)he will be noticed.

However, in the Juvenile Delinquent scenario, even successful attackers may have made use of different sets of actions, due to the three different ways to penetrate the house (\codefont{PenetrateHouse} \lil{-OR->} \codefont{\{OpenPassageDoor, EnterViaWindow, GarageAttack\}}). In fact, the average noticeability value of the \codefont{Best} attacker is just over~$3$, that of the \codefont{AverageA} %is close to~$6$, that of the 
and \codefont{Worst} attackers is just over~$9$, 
while also in this case the repeated attack attempts of the \codefont{AverageB} attacker guarantee that (s)he will be noticed.

\ourtool thus allows to analyze the risk of getting caught for different types of behavior of a concrete Cat Burglar or Juvenile Delinquent and to estimate who runs less risk. 
Secur\emph{IT}ree considers such explicit dynamic attack behavior in a slightly different way. It offers advanced analysis functionalities to estimate the risk of scenarios by combining the impact of attacks and the so-called capabilistic attack propensity, which is expressed by considering feasibility (e.g.\ cost or resources) vs.\ benefits (rewards) and detriments incurred in attacks.

%\begin{figure}[ht]
%\centering
%%\subfloat[Cat Burglar]{
%\includegraphics[width=0.45\columnwidth]
%{figures/CatBurglarplotSquare}
%%}
%%\subfloat[Juvenile Delinquent]{
%\quad
%\includegraphics[width=0.45\columnwidth]{figures/JuvenileDelinquentplotSquare}
%%}
%%\vspace{-0.35cm}
%\caption{\label{figure:plotCatBurglarJuvenile}Statistical analysis of noticeability on \lq\lq Cat Burglar\rq\rq\ (left) and \lq\lq Juvenile Delinquent\rq\rq\ (right)}
%\end{figure}

 %
 %\vspace*{-0.15cm}

\subsection{Case Study 3: Countermeasures}
\label{section:case3}
As in the previous sections, we focus on an influential approach to attack trees, \emph{attack countermeasure trees}~\cite{act}, which has inspired some of \ourtool's modeling features. 
We show how \ourtool DSL can specify the novel reactive defense mechanisms that were introduced in attack countermeasure trees, namely \emph{detection events\/} that model defensive mechanisms to detect that an attack is being attempted and \emph{measure events\/} that model defensive mechanisms to mitigate the effect of an attack. 

\begin{figure}[ht]
\centering
\includegraphics[width=\linewidth]{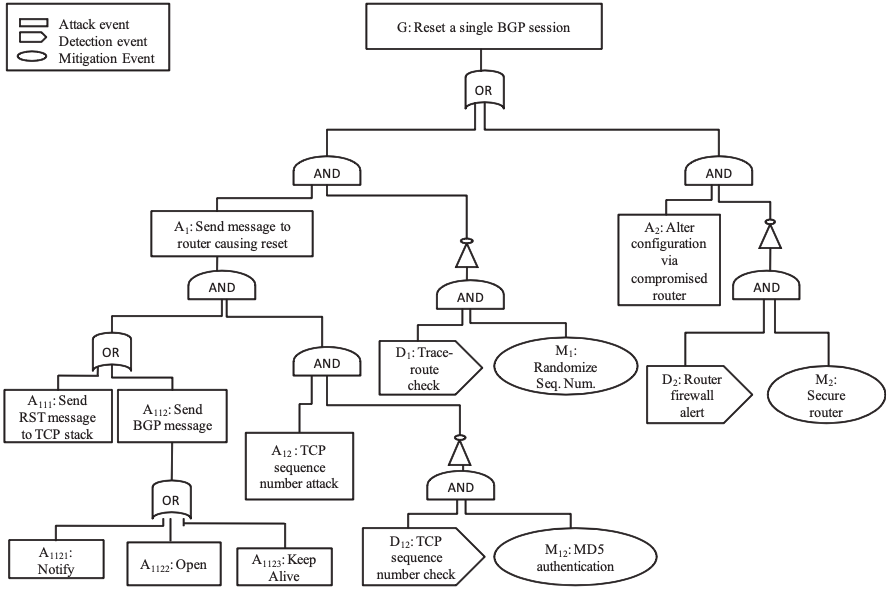}
%\vspace*{-0.25cm}
\caption{\label{figure:BGP}Countermeasure tree for \lq\lq Resetting BGP\rq\rq~\cite{act}}
\end{figure}

\subsubsection{Countermeasures Against \lq\lq Resetting BGP\rq\rq}

As illustrative example, we use a case study from~\cite{act}, namely an attack countermeasure tree modeling defensive mechanisms against resetting attacks on the so-called Border Gateway Protocol (BGP). Fig.~\ref{figure:BGP}, reproduced from~\cite{act}, depicts the attack countermeasure tree for this scenario. The idea is to model a known denial-of-service attack on the BGP: the attacker tries to reset a BGP session again and again to prevent communication. 
Such attacks consist of several steps, some of which can be detected and mitigated with well-known techniques (e.g.\ \codefont{TCP sequence num\-ber attack}s (\codefont{A12})  can be detected with \codefont{TCP sequen\-ce number check}s (\codefont{D12}), and a mitigation mechanism  for such attacks is using \codefont{MD5 authentication} (\codefont{M12})).

\subsubsection{Countermeasures in \ourtool{}}
\code{code:BGP} shows how to model the attack countermeasure tree of Fig.~\ref{figure:BGP} in \ourtool.

In particular, we remark the following: 
\begin{itemize}
\item
%(i)~
detection events \codefont{D12}, \codefont{D1} and \codefont{D2} are modeled as countermeasure nodes; the attacks \codefont{A12}, \codefont{A1} and \codefont{A2} they intend to detect, resp., are specified accordingly (cf.\ \lline{detection} in \code{code:BGP});
\item 
%(ii)~
measure events \codefont{M12}, \codefont{M1} and \codefont{DM2} are modeled as defense nodes (cf.\ \lline{measure} in \code{code:BGP});
the attacks \codefont{A12}, \codefont{A1} and \codefont{A2} they mitigate, resp., are specified as attack effectiveness block (cf.\ \lline{effective} in \code{code:BGP});
\item 
%and (iii)~
the relation between a detection event \codefont{D} and its triggered mitigation event \codefont{M} is modeled in \ourtool by specifying defense node \codefont{D} as a refinement of countermeasure node \codefont{M} (cf.\ \llines{defense1}{defensen} in \code{code:BGP}).
\end{itemize}

%caption={Attack tree of Fig.~\ref{figure:BGP} in \ourtool},
%\begin{wrapfigure}{l}{0.512\linewidth}
%\vspace{-0.1cm}
\begin{lstlisting}[float=t,language=spec,caption={Fig.~\ref{figure:BGP} in \ourtool},captionpos=b, label={code:BGP},numbers=left,escapeinside={@}{@}]
begin attack nodes
 G A1 A111 A112 A1121 A1122 A1123 A12 A2 OR1 	
end attack nodes

begin defense nodes @\label{measure}@
 M12 M1 M2
end defense nodes

begin countermeasure nodes @\label{detection}@
 D12 = {A12}, D1 = {A1}, D2 = {A2}
end countermeasure nodes

begin attack diagram
 G -OR-> {A1, A2} 
 A1 -AND-> {OR1, A12}
 OR1 -OR-> {A111, A112}
 A112 -OR->{A1121,A1122,A1123}
 D12 -AND-> {M12} @\label{defense1}@
 D1 -AND-> {M1}
 D2 -AND-> {M2}  @\label{defensen}@
end attack diagram

begin attack detection rates
 A1 = 0.5, A12 = 0.5, A2 = 0.5
end attack detection rates

begin defense effectiveness @\label{effective}@
 M12(ALL, A12) = 0.5, M1(ALL, A1) = 0.5, M2(ALL, A2) = 0.5
end defense effectiveness
\end{lstlisting}
%\vspace{-0.25cm}
%\end{wrapfigure}

\subsubsection{Complementing the Analysis of~\cite{act} with \ourtool{}}

The approach in~\cite{act} includes rich %a rich variety of 
analyses for attack countermeasure trees, including success probabilities, costs and impact of attacks and defensive mechanisms.  
\ourtool can augment such analyses with quantitative verification of specific attacker profiles. To illustrate this, we modeled three profiles:
\begin{description}[partopsep=-4pt,topsep=2pt,parsep=4pt,itemsep=-4pt]
\item{\codefont{Random}:} an attacker that randomly tries attacks until the main attack goal is achieved;
\item{\codefont{Noisy}:} like \codefont{Random} but tries attacks for which countermeasures exist with higher probability with respect to those for which no countermeasure exists;
\item{\codefont{Sneaky}:} like \codefont{Random} but tries attacks for which countermeasures exist with lower probability with respect to those for which no countermeasure exists.
\end{description}

\begin{figure}[h]%{r}{0.625\columnwidth}
\centering
\begin{overpic}[width=\columnwidth]{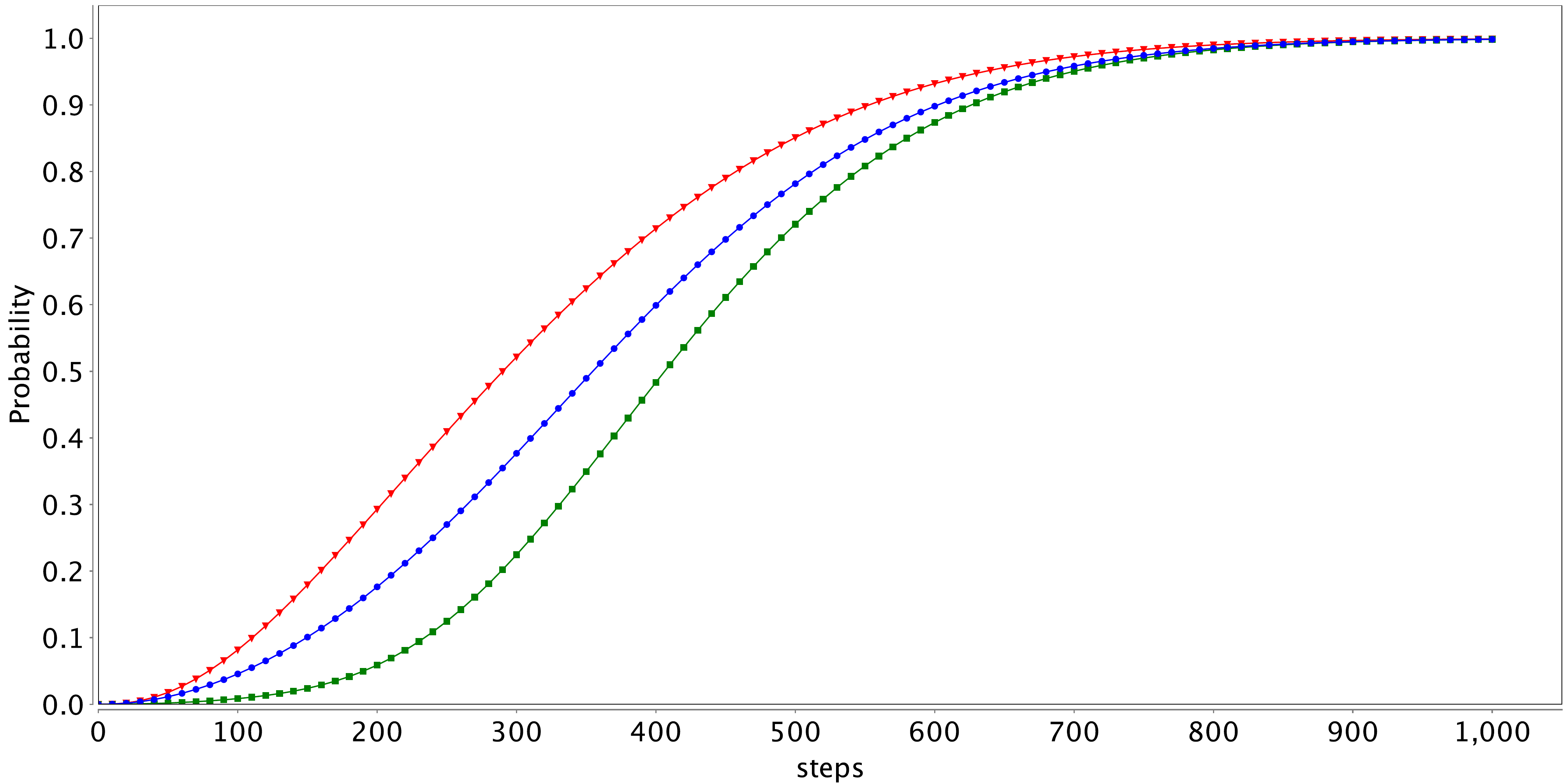}
     \put(80,10){\includegraphics[width=0.15\columnwidth]{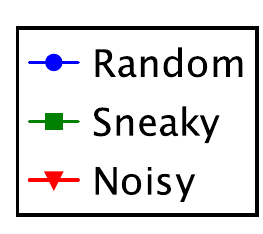}}  
  \end{overpic}
%\includegraphics[width=0.9\columnwidth]{figures/bgp-pri`sm-plot.pdf}
%\hspace{-2.0cm}
%\includegraphics[width=0.1\columnwidth]{figures/bgp-prism-legend.pdf}

%\includegraphics[width=0.7\columnwidth]{figures/BGPPlot}
%\includegraphics[width=0.6\columnwidth]{figures/Placeholder}
%\includegraphics[width=0.95\columnwidth]{figures/BGPplot-Random}\\
%\includegraphics[width=0.95\columnwidth]{figures/BGPplot-Sneaky}\\
%\includegraphics[width=0.95\columnwidth]{figures/BGPplot-Noisy}
%\vspace*{-0.25cm}
\caption{\label{figure:plotBGP}Exact PMC analysis on \lq\lq Resetting BGP\rq\rq}
\end{figure}

We analysed this scenario using the PMC functionalities of PRISM. Indeed, the DMTCs for the attackers %in this section 
could be %are small enough to be 
generated by \ourtool, and handled by PRISM. 
We \mbox{\lil{label}ed} \lil{with} \codefont{hasG} all states \lil{when} \lil{has}(\codefont{G})   was satisfied.
The property we studied is the probability of success at each step, suitably formulated in the property specification language of PRISM. 
Fig.~\ref{figure:plotBGP}, generated by PRISM, shows the results of the analyses: since all attackers are given the chance to try again and again, they are all eventually successful, but they differ with respect to the amount of time needed to succeed. 
Paradoxically, the \codefont{Noisy} attacker converges faster, which means that the detection and measure events are not as effective as they should be. 

%The full specification of the model is available at {\color{red}TBD}.

\section{Related Work}
\label{section:related}

There is a large body of related work. Throughout the paper, we indicated some sources of inspiration, like attack profiles specified as automata to describe possible attack steps and their costs~\cite{LWS14,HKKS16} and the attack detection rates~\cite{securitree}, ordered attacks~\cite{CY07} and countermeasures~\cite{act} treated in Sections~\ref{section:case1},~\ref{section:case2} and~\ref{section:case3}, resp.
A recent study by Wide\l{} et al.~\cite{WAFP19} classified existing approaches integrating attack tree-based modeling and formal methods along three dimensions. 
We believe that \ourtool can act as a unification of those dimensions. In this section, 
we detail the dimensions and relate \ourtool to existing approaches. 

The first dimension (a major focus of the large-scale EU project TRESPASS~\cite{TRES}) is the
generation of attack trees from scenarios. The main difficulty is to find a compact and effective representation, 
%of the tree, 
knowing that structurally different %attack 
trees can capture the same information. 
A representative contribution in this area is the ATSyRa toolset~\cite{PAV15}. %of Pinchinat et al.~\cite{PAV15}. 
An original and crucial feature of ATSyRa is the support for high-level actions 
(which can be seen as a sub-goal of the attacker) to specify how sequences of actions can be
abstracted and structured. Those high-level actions can later be used in a refactorization 
and hence better representation of the tree. The contribution is packed up in an elegant 
Eclipse plugin which makes it easily accessible to the uninitiated. Another contribution 
is the process-algebraic generation approach from Vigo and the Nielsons~\cite{VNN14}, 
where attacks are generated from flow constraints using a SAT solver, 
and a value-passing quality calculus is used to represent how an attacker can reach a given location. 
Our approach is not concerned with the synthesis of attack trees, but those techniques and tools could be combined in \ourtool, to complement them with analysis capabilities.
%For instance, ATSyRa could be combined in a complementary way: 
%\ourtool would benefit from both the Eclipse plugin and the scenario generation of ATSyRa, 
%while ATSyRa would benefit from all analysis primitives of \ourtool. 
%We believe this conclusion applies to many scenario generation tools.

The second dimension in~\cite{WAFP19} is that of giving a rigorous mathematical meaning 
to (extended) attack trees. The objective is to address a wide range of static problems, 
like comparing trees or enumerating the attacks. 
Well-known representations include Boolean function-based semantics, multisets, 
and linear logics (cf.~\cite{WAFP19,APK17} and their references). 
This research trend is very similar to the one applied to feature diagrams~\cite{CSW08}, 
and it is likely that many results from the software engineering community concerning product lines
or configurable systems can be transferred to the security domain~\cite{BLLV20}. 
It is worth noticing that the above mentioned approaches do not permit reasoning 
on the order of steps of the attack, a distinguishing feature that our approach has adopted, together with the ability to undo attack action. 
%Attack trees modeled in \ourtool directly translate to Boolean functions, and thus the comparison with this line of research is outside our scope.
 
More recently, several researchers have suggested to extend attack tree representations 
with their environment, %that of its environment, 
i.e.\ the attacker and the system under attack. For  example, the authors of~\cite{GHLLO16,HJLLP17} not only consider the attack tree itself, 
but also a transition system representation of an attacker model.
The separation of the attack tree from the attacker model as we do in \ourtool is fundamental to avoid confusion as explained by Mantel et al.~\cite{DBLP:conf/csfw/MantelP19}.  
This addition allows 
one to reason not only on static problems, but also on dynamic ones. For instance, 
one can make hypotheses on attack step sequences or extract correlations between
step orders. In addition, the use of transition system-based representations allows one to encompass a model of the system under attack, and by consequences of 
(the order of) its defenses~\cite{KMRS14}. In this context, contributions like~\cite{GHLLO16} 
consider that defenses are fixed a priori, while the game-based approach 
of Aslanyan et al.~\cite{ANP16} allows one to propose them dynamically to react 
to specific orders of attack steps. Observe that the latter proposal generalizes 
the sequential conjunction approach of Jhawar et al.~\cite{JKMRT15}. 
\ourtool follows the approach of Aslanyan et al., 
but uses SMC~\cite{LLTYSG19} (in addition to exact PMC), 
a simulation-based approach that is less precise but more effective than
the exhaustive state-space exploration of the game-based approach. 
Moreover, \ourtool offers a richer language to express constraints between attack steps 
and the behavior of the system under attack.

The third dimension proposed in~\cite{WAFP19} is that of adding quantitative algorithms 
to reason on (extensions of) attack trees. This is achieved by enriching attack trees 
with quantitative information, like the cost or probability of an attack step. 
In this context, static techniques can still be used to answer extended membership queries 
such as computing the cost of an attack, the Pareto optimal attack for two or more quantitative parameters, or the optimal countermeasures~\cite{AN15,DBLP:conf/csfw/FilaW19,FW20}. However, as observed by Kordy et al., 
minimal representations no longer exist~\cite{KMS12,KPS16}, which drastically complicates 
both the comparison and the synthesis of trees. Quantitative analysis extends to the dynamic case, 
meaning one can benefit from all the recent work on quantitative formal verification, 
where the attacker model can remain non-deterministic or even become stochastic. 
One can then synthesize strategies of the attacker that belongs to the tree and for which 
the cost is at most a certain value. Over the last five years, a wide range of
such techniques has been proposed. Some of those techniques were developed 
by Legay et al.~\cite{GHLLO16,HJLLP17}. %, already cited for the Boolean case. 
These approaches rely on a quantitative representation of the attacker together with a 
timed automata-based model for the system. Defenses are provided a priori. 
The approaches were implemented in the UPPAAL framework, 
which allows one to use extensions like UPPAAL SMC~\cite{DLLMP15} 
to compute the probability or cost of an attack.
In case non-determinism is added to the attacker model, 
UPPAAL Stratego~\cite{DJLMT15} can be used to synthesize strategies. 

\ourtool goes further than~\cite{GHLLO16,HJLLP17} by (i)~proposing a DSL and 
(ii)~allowing to not only quantify the number of attack steps, but also offering 
a rich process-algebraic language to impose conditions between steps as well as
between defenses that can moreover be added at runtime. However, \ourtool does 
not offer non-determinism for attackers. This may be needed to reason on the use of several strategies. 
A solution could be to add non-deterministic aspects to the DSL, 
and extend our DTMC exporter to an exporter for Markov decision processes, or in the input language of UPPAAL if also time aspects were to be considered, or to combine \ourtool's SMC engine with the Plasma Plugin for non-deterministic
systems~\cite{DLST15}. The approach by Aslanyan et al.~\cite{ANP16} allows reasoning on causality between steps and non-deterministic attackers, 
but restricted to Boolean causalities called waves, and without DSL. 
Stoelinga et al.\ also proposed dynamic approaches to analyze attack trees via SMC. 
Those approaches are covered and extended by \ourtool, 
especially concerning (i)~the causality part and (ii)~the DSL, 
which is restricted to the query part with LOCKS~\cite{KRS18}. 
Finally, compared to the three approaches mentioned above, 
only \ourtool proposes a fully dedicated and maintained open-source toolset.

\section{Conclusion and Future Work}
\label{section:future}

We instantiated \qflan in the quantitative security risk modeling and analysis domain, and applied the outcome, \ourtool, to 3 case studies %studied in
from well-known tools from the graph-based risk modeling and analysis domain. 
By enhancing the analysis features of % offered by 
these tools with either exact or statistical verification of probabilistic attack scenarios, \ourtool constitutes a significant contribution to the domain's toolsets. 

The generalization and subsequent instantiation of \qflan was feasible since it is open source, a distinguishing feature of \ourtool among the toolsets available in the domain. \ourtool's DTMC exporting facilities moreover permit tool-chaining with probabilistic model checkers for models of sizes that do not require SMC. 

\ourtool could be further enriched in several directions. 
First, we %currently 
propagate the value of an attribute of a %tree 
node as the sum of the attribute's values of its descendants. This could be generalized to attribute-specific formulae as in Secur\emph{IT}ree, in which, e.g., the noticeability value of a node with $n$ descendants $d_1, d_2, \ldots, d_n$ is computed as $1-((1-d_1)(1-d_2)\cdots(1-d_n))$.
Most properties analysed so far with \ourtool concern logical requirements. Recently, SMC has also been used to compare system behavior via simulation~\cite{LLMNN17}. We could compare the behavior of two attackers via simulation or their effect  on two different attack-defense diagrams. %Finally,  

We also plan to consider non-deterministic and game aspects along the lines of~\cite{ANP16,GHLLO16,HJLLP17}, as discussed in detail in Section~\ref{section:related}, as well as synthesis of attack profiles and countermeasures (cf., e.g.,~\cite{FW20}) for underspecified attack profiles.  
%and to synthesize %the
%attackers with the best %chance of success by %omitting some %of the 
%rates from the transitions %of the attack behavior model %and %rather 
%derive them.

\section*{Acknowledgment}
Supported by EU H2020 SU-ICT-03-2018 
project 830929 CyberSec4Europe.
%\newpage
%\clearpage

%\IEEEtriggeratref{54}
\bibliographystyle{IEEEtran}
\bibliography{ref} 

\providecommand{\noopsort}[1]{}
\begin{thebibliography}{10}
\providecommand{\url}[1]{#1}
\csname url@samestyle\endcsname
\providecommand{\newblock}{\relax}
\providecommand{\bibinfo}[2]{#2}
\providecommand{\BIBentrySTDinterwordspacing}{\spaceskip=0pt\relax}
\providecommand{\BIBentryALTinterwordstretchfactor}{4}
\providecommand{\BIBentryALTinterwordspacing}{\spaceskip=\fontdimen2\font plus
\BIBentryALTinterwordstretchfactor\fontdimen3\font minus
  \fontdimen4\font\relax}
\providecommand{\BIBforeignlanguage}[2]{{%
\expandafter\ifx\csname l@#1\endcsname\relax
\typeout{** WARNING: IEEEtran.bst: No hyphenation pattern has been}%
\typeout{** loaded for the language `#1'. Using the pattern for}%
\typeout{** the default language instead.}%
\else
\language=\csname l@#1\endcsname
\fi
#2}}
\providecommand{\BIBdecl}{\relax}
\BIBdecl

\bibitem{BDHHLLT12}
M.~Bozga, A.~David, A.~Hartmanns, H.~Hermanns, K.~G. Larsen, A.~Legay, and
  J.~Tretmans, ``{State-of-the-Art Tools and Techniques for Quantitative
  Modeling and Analysis of Embedded Systems},'' in \emph{Proceedings of the
  Conference on Design, Automation and Test in Europe (DATE'12)}.\hskip 1em
  plus 0.5em minus 0.4em\relax EDAA, 2012, pp. 370--375.

\bibitem{KL12}
J.~Katoen and K.~G. Larsen, ``{Quantitative Modelling and Analysis},'' in
  \emph{Proceedings of the 5th International Symposium on Leveraging
  Applications of Formal Methods, Verification and Validation: Applications and
  Case Studies (ISoLA'12)}, ser. Lecture Notes in Computer Science, T.~Margaria
  and B.~Steffen, Eds., vol. 7610.\hskip 1em plus 0.5em minus 0.4em\relax
  Springer, 2012, pp. 290--292.

\bibitem{HH15}
A.~Hartmanns and H.~Hermanns, ``In the quantitative automata zoo,'' \emph{Sci.
  Comput. Program.}, vol. 112, pp. 3--23, 2015.

\bibitem{KRS15}
R.~Kumar, E.~Ruijters, and M.~Stoelinga, ``{Quantitative Attack Tree Analysis
  via Priced Timed Automata},'' in \emph{Proceedings of the 13th International
  Conference on Formal Modeling and Analysis of Timed Systems (FORMATS'15)},
  ser. Lecture Notes in Computer Science, S.~Sankaranarayanan and E.~Vicario,
  Eds., vol. 9268.\hskip 1em plus 0.5em minus 0.4em\relax Springer, 2015, pp.
  156--171.

\bibitem{ANP16}
Z.~Aslanyan, F.~Nielson, and D.~Parker, ``{Quantitative Verification and
  Synthesis of Attack-Defence Scenarios},'' in \emph{Proceedings of the {IEEE}
  29th Computer Security Foundations Symposium (CSF'16)}.\hskip 1em plus 0.5em
  minus 0.4em\relax IEEE, 2016, pp. 105--119.

\bibitem{BDH16}
M.~Bernardo, R.~{De Nicola}, and J.~Hillston, Eds., \emph{Formal Methods for
  the Quantitative Evaluation of Collective Adaptive Systems}, ser. Lecture
  Notes in Computer Science, vol. 9700.\hskip 1em plus 0.5em minus 0.4em\relax
  Springer, 2016.

\bibitem{KS17}
R.~Kumar and M.~Stoelinga, ``{Quantitative Security and Safety Analysis with
  Attack-Fault Trees},'' in \emph{Proceedings of the 18th IEEE International
  Symposium on High Assurance Systems Engineering (HASE'17)}.\hskip 1em plus
  0.5em minus 0.4em\relax IEEE, 2017, pp. 25--32.

\bibitem{BLLV18}
M.~H. {\noopsort{Beek}}ter~Beek, A.~Legay, A.~{Lluch Lafuente}, and A.~Vandin,
  ``A framework for quantitative modeling and analysis of highly
  (re)configurable systems,'' \emph{IEEE Trans. Softw. Eng.}, vol.~46, no.~3,
  pp. 321--345, 2020.

\bibitem{BL19b}
M.~H. {\noopsort{Beek}}ter~Beek and A.~Legay, ``Quantitative variability
  modelling and analysis,'' \emph{Int. J. Softw. Tools Technol. Transf.},
  vol.~21, no.~6, pp. 607--612, 2019.

\bibitem{HHHKKKPQRS19}
E.~M. Hahn, A.~Hartmanns, C.~Hensel, M.~Klauck, J.~Klein,
  J.~Kret{\'{\i}}nsk{\'{y}}, D.~Parker, T.~Quatmann, E.~Ruijters, and
  M.~Steinmetz, ``The 2019 comparison of tools for the analysis of quantitative
  formal models,'' in \emph{Proceedings of the 25th International Conference on
  Tools and Algorithms for the Construction and Analysis of Systems: TOOLympics
  (TACAS'19)}, ser. Lecture Notes in Computer Science, D.~Beyer, M.~Huisman,
  F.~Kordon, and B.~Steffen, Eds., vol. 11429.\hskip 1em plus 0.5em minus
  0.4em\relax Springer, 2019, pp. 69--92.

\bibitem{Agha18}
G.~Agha and K.~Palmskog, ``A survey of statistical model checking,'' \emph{ACM
  Trans. Model. Comp. Simul.}, vol.~28, no.~1, pp. 6:1--6:39, 2018.

\bibitem{LLTYSG19}
A.~Legay, A.~Lukina, L.~Traonouez, J.~Yang, S.~A. Smolka, and R.~Grosu,
  ``{Statistical Model Checking},'' in \emph{Computing and Software Science:
  State of the Art and Perspectives}, ser. Lecture Notes in Computer Science,
  B.~Steffen and G.~J. Woeginger, Eds.\hskip 1em plus 0.5em minus 0.4em\relax
  Springer, 2019, vol. 10000, pp. 478--504.

\bibitem{LWS14}
A.~Lenin, J.~Willemson, and D.~P. Sari, ``{Attacker Profiling in Quantitative
  Security Assessment Based on Attack Trees},'' in \emph{Proceedings of the
  19th Nordic Conference on Secure IT Systems (NordSec'14)}, ser. Lecture Notes
  in Computer Science, K.~Bernsmed and S.~Fischer{-}H{\"{u}}bner, Eds., vol.
  8788.\hskip 1em plus 0.5em minus 0.4em\relax Springer, 2014, pp. 199--212.

\bibitem{GHLLO16}
O.~Gadyatskaya, R.~R. Hansen, K.~G. Larsen, A.~Legay, M.~C. Olesen, and D.~B.
  Poulsen, ``{Modelling Attack-defense Trees Using Timed Automata},'' in
  \emph{Proceedings of the 14th International Conference on Formal Modeling and
  Analysis of Timed Systems (FORMATS'16)}, ser. Lecture Notes in Computer
  Science, M.~Fr{\"{a}}nzle and N.~Markey, Eds., vol. 9884.\hskip 1em plus
  0.5em minus 0.4em\relax Springer, 2016, pp. 35--50.

\bibitem{HJLLP17}
R.~R. Hansen, P.~G. Jensen, K.~G. Larsen, A.~Legay, and D.~B. Poulsen,
  ``{Quantitative Evaluation of Attack Defense Trees Using Stochastic Timed
  Automata},'' in \emph{Proceedings of the 4th International Workshop on
  Graphical Models for Security (GraMSec'17)}, ser. Lecture Notes in Computer
  Science, P.~Liu, S.~Mauw, and K.~St{\o}len, Eds., vol. 10744.\hskip 1em plus
  0.5em minus 0.4em\relax Springer, 2017, pp. 75--90.

\bibitem{KSRYHBRS18}
R.~Kumar, S.~Schivo, E.~Ruijters, B.~M. Yildiz, D.~Huistra, J.~Brandt,
  A.~Rensink, and M.~Stoelinga, ``{Effective Analysis of Attack Trees: {A}
  Model-Driven Approach},'' in \emph{Proceedings of the 21st International
  Conference on Fundamental Approaches to Software Engineering (FASE'18)}, ser.
  Lecture Notes in Computer Science, A.~Russo and A.~Sch{\"{u}}rr, Eds., vol.
  10802.\hskip 1em plus 0.5em minus 0.4em\relax Springer, 2018, pp. 56--73.

\bibitem{CHVB18}
E.~M. Clarke, T.~A. Henzinger, H.~Veith, and R.~Bloem, Eds., \emph{Handbook of
  Model Checking}.\hskip 1em plus 0.5em minus 0.4em\relax Springer, 2018.

\bibitem{AT}
\BIBentryALTinterwordspacing
B.~Schneier, ``Attack trees,'' \emph{Dr. Dobb's Journal}, 1999. [Online].
  Available:
  \url{https://www.schneier.com/academic/archives/1999/12/attack_trees.html}
\BIBentrySTDinterwordspacing

\bibitem{foundationsAT}
S.~Mauw and M.~Oostdijk, ``{Foundations of Attack Trees},'' in
  \emph{Proceedings of the 8th International Conference on Information Security
  and Cryptology (ICISC'05)}, ser. Lecture Notes in Computer Science, D.~Won
  and S.~Kim, Eds., vol. 3935.\hskip 1em plus 0.5em minus 0.4em\relax Springer,
  2005, pp. 186--198.

\bibitem{foundationsADT}
B.~Kordy, S.~Mauw, S.~Radomirovi{\'{c}}, and P.~Schweitzer, ``{Foundations of
  Attack-Defense Trees},'' in \emph{Proceedings of the 7th International
  Workshop on Formal Aspects in Security and Trust (FAST'10)}, ser. Lecture
  Notes in Computer Science, P.~Degano, S.~Etalle, and J.~Guttman, Eds., vol.
  6561.\hskip 1em plus 0.5em minus 0.4em\relax Springer, 2011, pp. 80--95.

\bibitem{survey}
B.~Kordy, L.~Pi{\`{e}}tre{-}Cambac{\'{e}}d{\`{e}}s, and P.~Schweitzer,
  ``{{DAG}-based attack and defense modeling: {D}on't miss the forest for the
  attack trees},'' \emph{Comput. Sci. Rev.}, vol. 13--14, pp. 1--38, 2014.

\bibitem{HKCH17}
J.~B. Hong, D.~S. Kim, C.~Chung, and D.~Huang, ``{A survey on the usability and
  practical applications of Graphical Security Models},'' \emph{Comput. Sci.
  Rev.}, vol.~26, pp. 1--16, 2017.

\bibitem{WAFP19}
W.~Wide\l{}, M.~Audinot, B.~Fila, and S.~Pinchinat, ``{Beyond 2014: Formal
  Methods for Attack Tree--Based Security Modeling},'' \emph{ACM Comput.
  Surv.}, vol.~52, no.~4, pp. 75:1--75:36, 2019.

\bibitem{securitree}
\BIBentryALTinterwordspacing
{Amenaza Technologies Limited}, \emph{{The
  SecuITree\textsuperscript{\textregistered} BurgleHouse Tutorial (a.k.a., Who
  wants to be a Cat Burglar?)}}, 2nd~ed., 2006, (cf.~\cite{whitepaper}).
  [Online]. Available:
  \url{https://www.amenaza.com/downloads/docs/Tutorial.pdf}
\BIBentrySTDinterwordspacing

\bibitem{ADTool}
B.~Kordy, P.~Kordy, S.~Mauw, and P.~Schweitzer, ``{{ADTool}: Security Analysis
  with Attack--Defense Trees},'' in \emph{Proceedings of the 10th International
  Conference on Quantitative Evaluation of Systems (QEST'13)}, ser. Lecture
  Notes in Computer Science, K.~Joshi, M.~Siegle, M.~Stoelinga, and P.~R.
  D'Argenio, Eds., vol. 8054.\hskip 1em plus 0.5em minus 0.4em\relax Springer,
  2013, pp. 173--176.

\bibitem{KKB16}
B.~Kordy, P.~Kordy, and Y.~van~den Boom, ``{SPTool -- Equivalence Checker for
  SAND Attack Trees},'' in \emph{Proceedings of the 11th International
  Conference on Risks and Security of Internet and Systems (CRiSIS'16)}, ser.
  Lecture Notes in Computer Science, F.~Cuppens, N.~Cuppens, J.~Lanet, and
  A.~Legay, Eds., vol. 10158.\hskip 1em plus 0.5em minus 0.4em\relax Springer,
  2016, pp. 105--113.

\bibitem{SV13}
S.~Sebastio and A.~Vandin, ``{MultiVeStA: Statistical Model Checking for
  Discrete Event Simulators},'' in \emph{Proceedings of the 7th International
  Conference on Performance Evaluation Methodologies and Tools
  (ValueTools'13)}.\hskip 1em plus 0.5em minus 0.4em\relax ACM, 2013, pp.
  310--315.

\bibitem{GRV17}
S.~Gilmore, D.~Reijsbergen, and A.~Vandin, ``{Transient and Steady-State
  Statistical Analysis for Discrete Event Simulators},'' in \emph{Proceedings
  of the 13th International Conference on Integrated Formal Methods (IFM'17)},
  ser. Lecture Notes in Computer Science, N.~Polikarpova and S.~Schneider,
  Eds., vol. 10510.\hskip 1em plus 0.5em minus 0.4em\relax Springer, 2017, pp.
  145--160.

\bibitem{KNP11}
M.~Z. Kwiatkowska, G.~Norman, and D.~Parker, ``{{PRISM} 4.0: Verification of
  Probabilistic Real-Time Systems},'' in \emph{CAV}, ser. Lecture Notes in
  Computer Science, G.~Gopalakrishnan and S.~Qadeer, Eds., vol. 6806.\hskip 1em
  plus 0.5em minus 0.4em\relax Springer, 2011, pp. 585--591.

\bibitem{DJKV17}
C.~Dehnert, S.~Junges, J.~Katoen, and M.~Volk, ``{A Storm is Coming: A Modern
  Probabilistic Model Checker},'' in \emph{CAV}, ser. Lecture Notes in Computer
  Science, R.~Majumdar and V.~Kun{\v{c}}ak, Eds., vol. 10427.\hskip 1em plus
  0.5em minus 0.4em\relax Springer, 2017, pp. 592--600.

\bibitem{CY07}
S.~A. {\c{C}}amtepe and B.~Yener, ``Modeling and detection of complex
  attacks,'' in \emph{Proceedings of the 3rd International Conference on
  Security and Privacy in Communication Networks (SecureComm'07)}.\hskip 1em
  plus 0.5em minus 0.4em\relax IEEE, 2007, pp. 234--243.

\bibitem{improved}
W.~Lv and W.~Li, ``{Space Based Information System Security Risk Evaluation
  Based on Improved Attack Trees},'' in \emph{Proceedings of the 3rd
  International Conference on Multimedia Information Networking and Security
  (MINES'11)}.\hskip 1em plus 0.5em minus 0.4em\relax IEEE, 2011, pp. 480--483.

\bibitem{act}
A.~Roy, D.~S. Kim, and K.~S. Trivedi, ``Attack countermeasure trees ({ACT}):
  towards unifying the constructs of attack and defense trees,'' \emph{Secur.
  Commun. Netw.}, vol.~5, no.~8, pp. 929--943, 2012.

\bibitem{KMS12}
B.~Kordy, S.~Mauw, and P.~Schweitzer, ``{Quantitative Questions on
  Attack-Defense Trees},'' in \emph{Proceedings of the 15th International
  Conference on Information Security and Cryptology (ICISC'12)}, ser. Lecture
  Notes in Computer Science, T.~Kwon, M.~Lee, and D.~Kwon, Eds., vol.
  7839.\hskip 1em plus 0.5em minus 0.4em\relax Springer, 2012, pp. 49--64.

\bibitem{whitepaper}
\BIBentryALTinterwordspacing
T.~R. Ingoldsby, ``{Attack Tree-based Threat Risk Analysis},'' Amenaza
  Technologies Limited, Tech. Rep., October 2013. [Online]. Available:
  \url{https://www.amenaza.com/downloads/docs/AttackTreeThreatRiskAnalysis.pdf}
\BIBentrySTDinterwordspacing

\bibitem{VBLL18}
A.~Vandin, M.~H. ter Beek, A.~Legay, and A.~{Lluch Lafuente}, ``{{QFLan}: {A}
  Tool for the Quantitative Analysis of Highly Reconfigurable Systems},'' in
  \emph{Proceedings of the 22nd International Symposium on Formal Methods
  (FM'18)}, ser. Lecture Notes in Computer Science, K.~Havelund, J.~Peleska,
  B.~Roscoe, and E.~de~Vink, Eds., vol. 10951.\hskip 1em plus 0.5em minus
  0.4em\relax Springer, 2018, pp. 329--337.

\bibitem{BK08}
\BIBentryALTinterwordspacing
C.~Baier and J.-P. Katoen, \emph{Principles of Model Checking}.\hskip 1em plus
  0.5em minus 0.4em\relax The MIT Press, 2008. [Online]. Available:
  \url{http://mitpress.mit.edu/books/principles-model-checking}
\BIBentrySTDinterwordspacing

\bibitem{DBLP:conf/ifm/GilmoreTV14}
S.~Gilmore, M.~Tribastone, and A.~Vandin, ``{An Analysis Pathway for the
  Quantitative Evaluation of Public Transport Systems},'' in \emph{Proceedings
  of the 11th International Conference on Integrated Formal Methods (IFM'14)},
  ser. Lecture Notes in Computer Science, E.~Albert and E.~Sekerinski, Eds.,
  vol. 8739.\hskip 1em plus 0.5em minus 0.4em\relax Springer, 2014, pp. 71--86.

\bibitem{BLLV16}
M.~H. {\noopsort{Beek}}ter~Beek, A.~Legay, A.~{Lluch Lafuente}, and A.~Vandin,
  ``{Statistical Model Checking for Product Lines},'' in \emph{Proceedings of
  the 7th International Symposium on Leveraging Applications of Formal Methods,
  Verification and Validation: Foundational Techniques (ISoLA'16)}, ser.
  Lecture Notes in Computer Science, T.~Margaria and B.~Steffen, Eds., vol.
  9952.\hskip 1em plus 0.5em minus 0.4em\relax Springer, 2016, pp. 114--133.

\bibitem{Filipovikj2016SimulinkTU}
P.~Filipovikj, N.~Mahmud, R.~Marinescu, C.~Seceleanu, O.~Ljungkrantz, and
  H.~L{\"o}nn, ``{Simulink to UPPAAL Statistical Model Checker: Analyzing
  Automotive Industrial Systems},'' in \emph{FM}, ser. LNCS, J.~Fitzgerald,
  C.~Heitmeyer, S.~Gnesi, and A.~Philippou, Eds., vol. 9995.\hskip 1em plus
  0.5em minus 0.4em\relax Springer, 2016, pp. 748--756.

\bibitem{ABFMSLQE17}
A.~Arnold, M.~Baleani, A.~Ferrari, M.~Marazza, V.~Senni, A.~Legay, J.~Quilbeuf,
  and C.~Etzien, ``{An Application of {SMC} to continuous validation of
  heterogeneous systems},'' \emph{{EAI} Endorsed Trans. Indust. Netw. {\&}
  Intellig. Syst.}, vol.~4, no.~10, 2017.

\bibitem{BDG17}
D.~Basile, F.~{Di Giandomenico}, and S.~Gnesi, ``{Statistical Model Checking of
  an Energy-Saving Cyber-Physical System in the Railway Domain},'' in
  \emph{Proceedings 32nd Symposium on Applied Computing (SAC)}.\hskip 1em plus
  0.5em minus 0.4em\relax ACM, 2017, pp. 1356--1363.

\bibitem{CLSQTL17}
Q.~Cappart, C.~Limbr{\'{e}}e, P.~Schaus, J.~Quilbeuf, L.~Traonouez, and
  A.~Legay, ``{Verification of Interlocking Systems Using Statistical Model
  Checking},'' in \emph{Proceedings of the 18th International Symposium on High
  Assurance Systems Engineering (HASE'17)}.\hskip 1em plus 0.5em minus
  0.4em\relax IEEE, 2017, pp. 61--68.

\bibitem{BBC18}
D.~Basile, M.~H. ter Beek, and V.~Ciancia, ``{Statistical Model Checking of a
  Moving Block Railway Signalling Scenario with {\sc Uppaal} SMC},'' in
  \emph{ISoLA}, ser. LNCS, T.~Margaria and B.~Steffen, Eds., vol. 11245.\hskip
  1em plus 0.5em minus 0.4em\relax Springer, 2018, pp. 372--391.

\bibitem{PFG18}
S.~Puch, M.~Fr{\"{a}}nzle, and S.~Gerwinn, ``{Quantitative Risk Assessment of
  Safety-Critical Systems via Guided Simulation for Rare Events},'' in
  \emph{ISoLA}, ser. LNCS, T.~Margaria and B.~Steffen, Eds., vol. 11245.\hskip
  1em plus 0.5em minus 0.4em\relax Springer, 2018, pp. 305--321.

\bibitem{BADFL19}
R.~Bao, J.~C. Attiogb{\'{e}}, B.~Delahaye, P.~Fournier, and D.~Lime,
  ``{Parametric Statistical Model Checking of {UAV} Flight Plan},'' in
  \emph{FORTE}, ser. LNCS, J.~A. P{\'{e}}rez and N.~Yoshida, Eds., vol.
  11535.\hskip 1em plus 0.5em minus 0.4em\relax Springer, 2019, pp. 57--74.

\bibitem{BBFL19}
D.~Basile, M.~H. ter Beek, A.~Ferrari, and A.~Legay, ``{Modelling and Analysing
  ERTMS L3 Moving Block Railway Signalling with Simulink and UPPAAL SMC},'' in
  \emph{FMICS}, ser. LNCS, K.~G. Larsen and T.~Willemse, Eds., vol.
  11687.\hskip 1em plus 0.5em minus 0.4em\relax Springer, 2019, pp. 1--21.

\bibitem{DBLP:conf/rssrail/BasileFRM19}
D.~Basile, A.~Fantechi, L.~Rucher, and G.~Mand{\`{o}}, ``Statistical model
  checking of hazards in an autonomous tramway positioning system,'' in
  \emph{RSSRail}, ser. LNCS, S.~Collart-Dutilleul, T.~Lecomte, and A.~B.
  Romanovsky, Eds., vol. 11495.\hskip 1em plus 0.5em minus 0.4em\relax
  Springer, 2019, pp. 41--58.

\bibitem{FMBBF20}
A.~Ferrari, F.~Mazzanti, D.~Basile, M.~H. ter Beek, and A.~Fantechi,
  ``{Comparing Formal Tools for System Design: a Judgment Study},'' in
  \emph{Proceedings 42nd International Conference on Software Engineering
  (ICSE)}.\hskip 1em plus 0.5em minus 0.4em\relax ACM, 2020.

\bibitem{GBP20}
H.~Garavel, M.~H. ter Beek, and J.~van~de Pol, ``{The 2020 Expert Survey on
  Formal Methods},'' in \emph{FMICS}, ser. LNCS, M.~ter Beek and
  D.~Ni{\v{c}}kovi{\'{c}}, Eds., vol. 12327.\hskip 1em plus 0.5em minus
  0.4em\relax Springer, 2020, pp. 3--69.

\bibitem{HKKS16}
H.~Hermanns, J.~Kr{\"{a}}mer, J.~Krc{\'{a}}l, and M.~Stoelinga, ``{The Value of
  Attack-Defence Diagrams},'' in \emph{Proceedings of the 5th International
  Conference on Principles of Security and Trust (POST'16)}, ser. Lecture Notes
  in Computer Science, F.~Piessens and L.~Vigan{\`{o}}, Eds., vol. 9635.\hskip
  1em plus 0.5em minus 0.4em\relax Springer, 2016, pp. 163--185.

\bibitem{TRES}
\BIBentryALTinterwordspacing
``{H2020 project on robusT Risk basEd Screening and alert System for PASSengers
  and luggage}.'' [Online]. Available:
  \url{https://www.tresspass.eu/The-project}
\BIBentrySTDinterwordspacing

\bibitem{PAV15}
S.~Pinchinat, M.~Acher, and D.~Vojtisek, ``{ATSyRa: An Integrated Environment
  for Synthesizing Attack Trees - (Tool Paper)},'' in \emph{Proceedings of the
  2nd International Workshop on Graphical Models for Security (GraMSec'15)},
  ser. Lecture Notes in Computer Science, S.~Mauw, B.~Kordy, and S.~Jajodia,
  Eds., vol. 9390.\hskip 1em plus 0.5em minus 0.4em\relax Springer, 2015, pp.
  97--101.

\bibitem{VNN14}
R.~Vigo, F.~Nielson, and H.~R. Nielson, ``Automated generation of attack
  trees,'' in \emph{Proceedings of the 27th {IEEE} Computer Security
  Foundations Symposium (CSF'14)}.\hskip 1em plus 0.5em minus 0.4em\relax IEEE,
  2014, pp. 337--350.

\bibitem{APK17}
M.~Audinot, S.~Pinchinat, and B.~Kordy, ``{Is My Attack Tree Correct?}'' in
  \emph{Proceedings 22nd European Symposium on Research in Computer Security
  (ESORICS'17)}, ser. Lecture Notes in Computer Science, S.~N. Foley,
  D.~Gollmann, and E.~Snekkenes, Eds., vol. 10492.\hskip 1em plus 0.5em minus
  0.4em\relax Springer, 2017, pp. 83--102.

\bibitem{CSW08}
K.~Czarnecki, S.~She, and A.~Wasowski, ``{Sample Spaces and Feature Models:
  There and Back Again},'' in \emph{Proceedings of the 12th International
  Software Product Lines Conference (SPLC'08)}.\hskip 1em plus 0.5em minus
  0.4em\relax IEEE, 2008, pp. 22--31.

\bibitem{BLLV20}
M.~H. {\noopsort{Beek}}ter~Beek, A.~Legay, A.~{Lluch Lafuente}, and A.~Vandin,
  ``{Variability meets Security},'' in \emph{Proceedings of the 14th
  International Working Conference on Variability Modelling of
  Software-intensive Systems (VaMoS'20)}.\hskip 1em plus 0.5em minus
  0.4em\relax ACM, 2020, pp. 11:1--11:9.

\bibitem{DBLP:conf/csfw/MantelP19}
H.~Mantel and C.~W. Probst, ``{On the Meaning and Purpose of Attack Trees},''
  in \emph{Proceedings of the 32nd {IEEE} Computer Security Foundations
  Symposium (CSF'19)}.\hskip 1em plus 0.5em minus 0.4em\relax IEEE, 2019, pp.
  184--199.

\bibitem{KMRS14}
B.~Kordy, S.~Mauw, S.~Radomirovic, and P.~Schweitzer, ``Attack-defense trees,''
  \emph{J. Log. Comput.}, vol.~24, no.~1, pp. 55--87, 2014.

\bibitem{JKMRT15}
R.~Jhawar, B.~Kordy, S.~Mauw, S.~Radomirovic, and R.~Trujillo{-}Rasua,
  ``{Attack Trees with Sequential Conjunction},'' in \emph{Proceedings of the
  30th {IFIP} {TC} 11 International Conference on {ICT} Systems Security and
  Privacy Protection (SEC'15)}, ser. {IFIP} Advances in Information and
  Communication Technology, H.~Federrath and D.~Gollmann, Eds., vol. 455.\hskip
  1em plus 0.5em minus 0.4em\relax Springer, 2015, pp. 339--353.

\bibitem{AN15}
Z.~Aslanyan and F.~Nielson, ``{Pareto Efficient Solutions of Attack-Defence
  Trees},'' in \emph{Proceedings of the 4th International Conference on
  Principles of Security and Trust (POST'15)}, ser. Lecture Notes in Computer
  Science, R.~Focardi and A.~C. Myers, Eds., vol. 9036.\hskip 1em plus 0.5em
  minus 0.4em\relax Springer, 2015, pp. 95--114.

\bibitem{DBLP:conf/csfw/FilaW19}
B.~Fila and W.~Wide\l{}, ``{Efficient Attack-Defense Tree Analysis using Pareto
  Attribute Domains},'' in \emph{Proceedings of the 32nd {IEEE} Computer
  Security Foundations Symposium (CSF'19)}.\hskip 1em plus 0.5em minus
  0.4em\relax IEEE, 2019, pp. 200--215.

\bibitem{FW20}
------, ``Exploiting attack-defense trees to find an optimal set of
  countermeasures,'' in \emph{Proceedings of the 33rd {IEEE} Computer Security
  Foundations Symposium (CSF'20)}.\hskip 1em plus 0.5em minus 0.4em\relax IEEE,
  2020, pp. 395--410.

\bibitem{KPS16}
B.~Kordy, M.~Pouly, and P.~Schweitzer, ``Probabilistic reasoning with graphical
  security models,'' \emph{Inf. Sci.}, vol. 342, pp. 111--131, 2016.

\bibitem{DLLMP15}
A.~David, K.~G. Larsen, A.~Legay, M.~Mikucionis, and D.~B. Poulsen,
  ``{\textsc{Uppaal} {SMC} tutorial},'' \emph{Int. J. Softw. Tools Technol.
  Transf.}, vol.~17, no.~4, pp. 397--415, 2015.

\bibitem{DJLMT15}
A.~David, P.~G. Jensen, K.~G. Larsen, M.~Miku{\v{c}}ionis, and J.~H. Taankvist,
  ``{Uppaal Stratego},'' in \emph{Proceedings of the 21st International
  Conference on Tools and Algorithms for the Construction and Analysis of
  Systems (TACAS'15)}, ser. Lecture Notes in Computer Science, C.~Baier and
  C.~Tinelli, Eds., vol. 9035.\hskip 1em plus 0.5em minus 0.4em\relax Springer,
  2015, pp. 206--211.

\bibitem{DLST15}
P.~D'Argenio, A.~Legay, S.~Sedwards, and L.~Traonouez, ``{Smart sampling for
  lightweight verification of Markov decision processes},'' \emph{Int. J.
  Softw. Tools Technol. Transf.}, vol.~17, no.~4, pp. 469--484, 2015.

\bibitem{KRS18}
R.~Kumar, A.~Rensink, and M.~Stoelinga, ``{{LOCKS}: a property specification
  language for security goals},'' in \emph{Proceedings of the 33rd Annual {ACM}
  Symposium on Applied Computing (SAC'18)}.\hskip 1em plus 0.5em minus
  0.4em\relax ACM, 2018, pp. 1907--1915.

\bibitem{LLMNN17}
K.~G. Larsen, A.~Legay, M.~Miku{\v{c}}ionis, B.~Nielsen, and U.~Nyman,
  ``{Compositional Testing of Real-Time Systems},'' in \emph{ModelEd, TestEd,
  TrustEd}, ser. Lecture Notes in Computer Science, J.~Katoen, R.~Langerak, and
  A.~Rensink, Eds.\hskip 1em plus 0.5em minus 0.4em\relax Springer, 2017, vol.
  10500, pp. 107--124.

\end{thebibliography}

\end{document}